\documentclass{aa} 
\usepackage{epsfig,times} 
\def\mincir{\raise -2.truept\hbox{\rlap{\hbox{$\sim$}}\raise5.truept \hbox{$<$}\ }} 
\def\mincireq{\hbox{\raise0.5ex\hbox{$<\lower1.06ex\hbox{$\kern-1.07em{\sim}$}$}}} 
\def\magcir{\raise-2.truept\hbox{\rlap{\hbox{$\sim$}}\raise5.truept \hbox{$>$}\ }} 
\def\gr{\kern 2pt\hbox{}^\circ{\kern -2pt K}} 

\def\G{\Gamma}

\def\_{\thinspace}

\def\cm{\thinspace{\rm cm}} 
\def\cm2{\thinspace{\rm cm}^2}

\def\be{\begin{equation}} 
\def\ee{\end{equation}}

\begin{document} 
 
\title{2-10 keV luminosity of high-mass binaries as a gauge of ongoing star-formation rate}                                   

\author{M. Persic\inst{1}, 
	Y. Rephaeli\inst{2,3},
	V. Braito\inst{4},
	M. Cappi\inst{5},
	R. Della Ceca\inst{4}, 
	A. Franceschini\inst{6}, and
	D. E. Gruber\inst{7}} 
 
\offprints{M.P.; e-mail: {\tt persic@ts.astro.it}} 
 
\institute{ 
INAF/Osservatorio Astronomico di Trieste, via G.B.Tiepolo 11, 34131 Trieste, Italy
	\and  
School of Physics and Astronomy, Tel Aviv University, Tel Aviv 69978, 
Israel
	\and
CASS, University of California, San Diego, La Jolla, CA 92093, USA
	\and
INAF/Osservatorio Astronomico di Brera, via Brera 28, 20121 Milano, Italy
	\and
INAF/IASF, via P.Gobetti 101, 40129 Bologna, Italy
	\and
Dipartimento di Astronomia, Universit\`a di Padova, vicolo Osservatorio 2, 35122 Padova, Italy
	\and
4789 Panorama Drive, San Diego, CA 92116, USA
}
\date{Received ..................; accepted ...................}

\abstract{ 
Based on recent work on spectral decomposition of the emission of star-forming 
galaxies, we assess whether the integrated 2-10 keV emission from high-mass X-ray 
binaries (HMXBs), $L_{2-10}^{\rm HMXB}$, can be used as a reliable estimator of 
ongoing star formation rate (SFR). Using a sample of 46 local ($z \mincir 0.1$) 
star-forming galaxies, and spectral modeling of {\it ASCA}, {\it BeppoSAX}, and 
{\it XMM}-Newton data, we demonstrate the existence of a linear SFR--$L_{2-10}^
{\rm HMXB}$ relation which holds over $\sim$5 decades in X-ray luminosity and SFR. 
The total 2-10 keV luminosity is {\it not} a precise SFR indicator because at low 
SFR (i.e., in normal and moderately-starbursting galaxies) it is substantially 
affected by the emission of low-mass X-ray binaries, which do not trace the current 
SFR due to their long evolution lifetimes, while at very high SFR (i.e., for very 
luminous FIR-selected galaxies) it is frequently affected by the presence of 
strongly obscured AGNs. The availability of purely SB-powered galaxies -- whose 
2-10 keV emission is mainly due to HMXBs -- allows us to properly calibrate the 
SFR--$L_{2-10}^{\rm HMXB}$ relation. The SFR--$L_{2-10}^{\rm HMXB}$ relation holds 
also for distant ($z \sim 1$) galaxies in the {\it Hubble} Deep Field North sample, 
for which we lack spectral information, but whose SFR can be estimated from deep 
radio data. If confirmed by more detailed observations, it may be possible to use 
the deduced relation to identify distant galaxies that are X-ray overluminous for 
their (independently estimated) SFR, and are therefore likely to hide strongly 
absorbed AGNs. 
}

\maketitle 
\markboth{Persic et al.: 2-10 keV HMXB luminosity as galactic SFR indicator}{} 
 
\section{Introduction}

X-ray emission of star-forming galaxies (SFGs) consists of various components including 
discrete sources, such as X-ray binaries and supernova remnants (SNRs), diffuse hot gas, 
Compton scattering of ambient FIR photons, and possibly an active nucleus. The resulting 
integrated spectra harbor the signatures of these emission components.

\begin{figure}
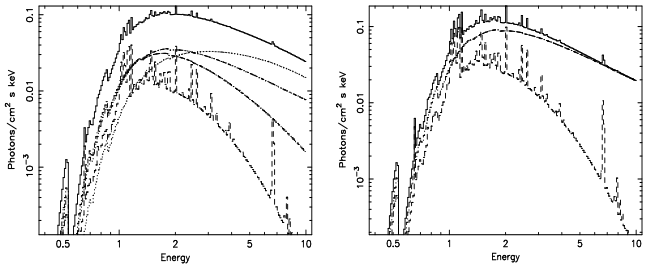

\vspace{3.4cm}
\includegraphics{0500fig1a.ps}
\hspace{.5cm} 
\includegraphics{0500fig1b.ps}

\caption{The integrated "stellar" emission in the template X-ray spectrum of a star-forming 
galaxy proposed by Persic \& Rephaeli (2002). Two different cases are shown: "standard" 
({\it left}) and "top-heavy" ({\it right}). Shown for the standard case in ascending order 
at 7 keV: SNRs, faint LMXBs, HMXBs, and bright LMXBs. On the right, in ascending order at 3 
keV, only SNRs and HMXBs are shown. The assumed 0.5-50 keV luminosities and number abundances 
are: log$L_{\rm x} = 37.7$ for HMXBs (50 objects) and high-luminosity LMXBs (70 objects), 
37.0 for SNRs (20 objects), and 36.7 for low-luminosity LMXBs (130 objects). The spectral 
components are normalized in energy flux in the 0.5-50 keV band. The spectrum is absorbed 
through a HI column density $N_{\rm H}= 10^{22}$ cm$^{-2}$.}
\end{figure}

Persic \& Rephaeli (2002, hereafter PR02) have quantitatively assessed the roles of the 
various X-ray emission mechanisms in SFGs. They have used an equilibrium stellar-population 
synthesis model of the Galactic population of X-ray binaries (Iben et al. 1995a,b) to deduce 
birthrates for interacting binaries; these, combined with estimates of the duration of the 
X-ray bright phase, have allowed PR02 to make realistic estimates of the relative (Galactic) 
abundances of high-mass and low-mass X-ray binaries (HMXBs, LMXBs). The abundance of SNRs 
(both Type II and Ia explosions) was also consistently estimated. From the literature PR02 
derived typical spectra for these classes of source. The spectral properties and relative 
abundances of the various classes of stellar sources determine the composite X-ray spectrum 
arising from a stellar population of Galactic composition. Therefore, the PR02 "stellar" 
contribution has then no essential degrees of freedom: fixed by the synthetic model of the 
Galactic population of binaries and by the observed X-ray spectra of the contributing 
components, it represents the X-ray spectrum emitted by a Galactic mix of HMXBs and LMXBs. 
As such, the stellar component of the PR02 template spectrum is likely to be appropriate for 
a quietly star forming galaxy like the Milky Way (see Fig.1-{\it left}). 
	\footnote{The diffuse non-stellar part of the PR02 synthetic spectrum has both 
	a thermal and a non-thermal component. The former is related to the SN-powered 
	outgoing galactic wind and is mostly relevant at energies $\mincir$1 keV. The 
	latter is due to Compton scattering of the SN-accelerated, radio-emitting 
	relativistic electrons off the FIR and CMB radiation fields, and it arguably 
	dominates the spectrum of SFGs at energies of $\magcir$30 keV (see Persic \& 
	Rephaeli 2002, 2003). }		

The PR02 approach has the flexibility of handling also the extreme cases of 'no ongoing SF' 
and 'very high ongoing SF' by switching off the HMXB and SNR components and, respectively, 
the LMXB component (see Fig.1-{\it right}). By letting the amplitudes of the various spectral 
components vary (while keeping their profiles fixed) in the spectral fit, one can use the 
PR02 procedure to determine the SF state of a given galaxy. Thus, this treatment is generally 
valid and not limited to any particular regime of SF activity. 

Based on their survey of galactic X-ray emission mechanisms, PR02 concluded that in the 2-10 keV 
energy range X-ray binaries of both types, HMXBs and LMXBs, are the most prominent components with 
the required spectral shapes (see Fig.1-{\it left}). These have either a power-law (PL) form for 
HMXBs, or cut-off PL for LMXBs, with observed ranges of spectral parameters (HMXBs: $\G \simeq 
1.0-1.4$, see White et al. 1983; LMXBs: $\G \simeq 0-1.4$, $E_{\rm c} \sim 5-10$ keV, see Christian 
\& Swank 1997) that provide good fits to the spatially unresolved {\it BeppoSAX} spectra of the 
most extensively observed nearby starburst galaxies (SBGs), M~82 and NGC~253 (see PR02), and to the 
{\it XMM-Newton} spectra of a number of distant Ultra-Luminous Infra-Red Galaxies (ULIRGs; 
Franceschini et al. 2003).

From a broader perspective, the template spectrum of PR02 provides a framework 
that may prove especially useful for interpreting low spatial resolution data on SFGs, 
either distant (from {\it Chandra} and {\it XMM}: e.g., Hornschemeier et al. 2001, 2003; 
Alexander et al. 2002; Bauer et al. 2002; Franceschini et al. 2003) or nearby (from 
{\it ASCA} and {\it BeppoSAX}: e.g., Cappi et al. 1999; Dahlem et al. 1998; Della Ceca 
1996, 1997, 1999; Moran et al. 1999). In particular, this template spectrum would allow 
the general possibility of measuring the {\it ongoing} star formation rate (SFR) in 
galaxies from their X-ray spectra, or perhaps -- for some galaxies --  directly from their 
X-ray luminosities. The basic notion is that ongoing SFR in a galaxy can be measured based 
on stellar X-ray sources which are both sufficiently bright for their collective emission 
to be unambiguously identified, and sufficiently short-lived so that they trace the 
'instantaneous' SFR. Of the three main types of stellar galactic X-ray sources, 

\noindent
{\it (i)} SNRs (which are X-ray bright over timescales $t_{\rm x} \sim 10^3$ yr) are 
the evolutionary outcome of massive ($8 \mincir M/M_\odot \mincir 40$) stars that explode 
on timescales ($5 \mincir \tau_{\rm ev}/{\rm Myr} \mincir 50$, see Maeder \& Meynet 1989) 
short compared with a typical SB duration ($\tau_{\rm SB} \mincir 100$ Myr). Hence SNRs do 
trace the istantaneous SFR; however, their relative emission in integrated SFG spectra is 
quite modest and hard to identify; 

\noindent
{\it (ii)} LMXBs ($t_{\rm x} \sim 10^7$ yr) do contribute significantly to the X-ray emission but, 
due to the long delay between their formation and the onset of their X-ray emission (the donor star 
has $M \mincir M_\odot$), they do not trace the current SFR; finally, 

\noindent
{\it (iii)} HMXBs ($t_{\rm x} \sim 10^4$ yr) provide a suitable combination of short 
delay between binary formation and onset of X-ray emission (the donor star has 
$M \magcir 8\,M_\odot$) and significant -- sometimes dominant -- relative X-ray emission. 

\noindent
Consequently, the measurement of ongoing SFR in galaxies hinges on our ability to separate 
out the HMXB contribution to the 2-10 keV luminosity, $L_{2-10}$. 

In this paper we extend the work of PR02 by examining ways in which $L_{2-10}$ can be 
used as an astrophysically motivated SFR estimator for SFGs. The possibility of using 
$L_{2-10}$ as an integral measure of the SFR has already been suggested (e.g.: Bauer et 
al. 2002; Grimm et al. 2003; Ranalli et al. 2003; Gilfanov et al. 2004). In this paper 
we propose that by using the HMXB portion of the hard X-ray luminosity, the SFR--$L_{\rm 
2-10}^{\rm HMXB}$ relation is universal and extends from normal galaxies (low SFR) to very 
actively starbursting galaxies (very high SFR). Given the ways in which $L_{\rm 2-10}$ 
can be contaminated, i.e. systematically from LMXBs at low-SFR regimes (PR02), and 
occasionally but quite frequently from AGNs at high-SFR regimes (Franceschini et al. 
2003), our results will emphasize the potential and the limitations of using -- 
over $\sim$5 decades in X-ray luminosity -- the 2-10 keV luminosity as an independent 
gauge of SFR in SFGs.

The paper is organized as follows. In section 2 we discuss galactic SFR indicators 
linked to the presence of short-lived massive ($\magcir 8 M_\odot$) main-sequence 
stars. Section 3 describes the sample of SFGs, which comprises objects with SBs of 
various strengths. In section 4 we discuss the effectiveness of the 2-10 keV luminosity 
of HMXBs as a SFR indicator. Section 5 summarizes our main results. The conclusions are 
in section 6. 

\begin{table*} 
\caption[] {Data I: The sample of ultra-luminous IR galaxies (ULIRGs)$^{(a)}$.} 
\begin{flushleft} 
\begin{tabular}{ l l l l l l l l}
\noalign{\smallskip} 
\hline 
\hline 
\noalign{\smallskip} 
Object         & D$^{(b)}$  & $f_{0.5-2}$               &  $f_{2-10}$              & Instr. & $f_{60}$ & $f_{100}$  & Notes$^{(c)}$     \\
               & [Mpc]      &[$10^{-14}$ erg s$^{-1}$]  &[$10^{-14}$ erg s$^{-1}$] &        &   [Jy]   &  [Jy]      &                   \\
\noalign{\smallskip}
\hline
\noalign{\smallskip}
IRAS~05189-2524&  170$^{[7]}$  &     ....          & 360.0$^{[7]}$     & SAX   &  13.25$^{[8]}$ &    11.84$^{[8]}$ & AGN$^{[7]}$        \\
IRAS~12112+0305&  291$^{[3]}$  &   1.53$^{[2]}$    &   1.62$^{[2]}$    & XMM   &  8.18$^{[3]}$  &    9.46$^{[3]}$  & SB                 \\
Mkn~231        &  169$^{[3]}$  &  29.06$^{[2,11]}$ &  84.54$^{[2,11]}$ & XMM   &   32.0$^{[3]}$ &    30.3$^{[3]}$  & SB + AGN$^{[2]}$   \\
Mkn~273        &  151$^{[3]}$  &  17.00$^{[4]}$    &  70.0$^{[4]}$     & ASCA  &  21.7$^{[3]}$  &    21.4$^{[3]}$  & SB + AGN$^{[4]}$   \\
IRAS~14348-1447&  330$^{[3]}$  &   3.08$^{[2]}$    &   1.91$^{[2]}$    & XMM   &  6.82$^{[3]}$  &    7.31$^{[3]}$  & SB                 \\
IRAS~15250-3609&  213$^{[3]}$  &   2.31$^{[2]}$    &   2.31$^{[2]}$    & XMM   &  7.10$^{[3]}$  &    5.93$^{[3]}$  & SB                 \\
Arp~220        &   73$^{[3]}$  &   8.0$^{[6]}$     &  18.0$^{[6]}$     & SAX   & 104.0$^{[3]}$  &   112.0$^{[3]}$  & SB                 \\
NGC~6240       &   97$^{[3]}$  &  64.0$^{[5]}$     & 190.0$^{[5,9]}$   & ASCA, SAX & 22.94$^{[3]}$ &   26.49$^{[3]}$& AGN$^{[9]}$       \\
IRAS~17208-0014&  170$^{[3]}$  &   7.56$^{[2]}$    &   4.54$^{[2]}$    & XMM   &  9.53$^{[3]}$  &   11.05$^{[3]}$  & SB                 \\
IRAS~19254-7245&  246$^{[3]}$  &  10.91$^{[2,10]}$ &  17.58$^{[2,10]}$ & XMM   &  5.5$^{[3]}$   &     5.8$^{[3]}$  & SB + AGN$^{[2]}$   \\
               &               &                   &                   &       &                &                  & (The Superantennae)\\
IRAS~20100-4156&  517$^{[3]}$  &   1.68$^{[2]}$    &   2.24$^{[2]}$    & XMM   &  5.2$^{[3]}$   &     5.2$^{[3]}$  & SB                 \\
IRAS~20551-4250&  171$^{[3]}$  &  52.93$^{[2]}$    &  99.56$^{[2]}$    & XMM   & 12.8$^{[3]}$   &    10.0$^{[3]}$  & SB + AGN$^{[2]}$   \\
IRAS~22491-1808&  309$^{[3]}$  &   0.61$^{[2]}$    &   0.65$^{[2]}$    & XMM   &  5.54$^{[3]}$  &    4.64$^{[3]}$  & SB                 \\
IRAS~23060+0505&  692$^{[3]}$  & 270.00$^{[1]}$    & 250.0$^{[1]}$     & ASCA  &  1.2$^{[3]}$   &     0.8$^{[3]}$  & AGN$^{[1]}$        \\
IRAS~23128-5919&  178$^{[3]}$  &  12.04$^{[2]}$    &  21.66$^{[2]}$    & XMM   &  10.8$^{[3]}$  &    11.0$^{[3]}$  & SB + AGN$^{[2]}$   \\

\noalign{\smallskip}
\hline
\hline
\end{tabular}
\end{flushleft}
\smallskip

$^{(a)}$ References: [1] Brandt et al. 1997; [2] Franceschini et al. 2003; [3] 
Genzel et al. 1998; [4] Iwasawa 1999; [5] Iwasawa \& Comastri 1998; [6] Iwasawa 
et al. 2001; [7] Severgnini et al. 2001; [8] Sanders et al. 2003; [9] Vignati
et al. 1999; [10] Braito et al. 2003; [11] Braito et al. 2004.

$^{(b)}$ Distances are taken from Genzel et al. (1998), or are computed consistently 
assuming H$_0=75$ km s$^{-1}$ Mpc$^{-1}$, $q_0=0.5$ otherwise.

$^{(c)}$ Component(s) dominating the 2-10 keV emission; other name(s).

\end{table*}

\begin{table*}
\caption[] {
Data II: 
The sample of local normal and starburst galaxies (SBG 
sample)$^{(a)}$.}
\begin{flushleft}
\begin{tabular}{ l  l  l  l  l  l  l  l  l  }
\noalign{\smallskip}
\hline
\hline
\noalign{\smallskip}
Object$^{(b)}$  &   D$^{(c)}$     & $f_{0.5-2}^{(d)}$ &  $f_{2-10}$       & Instr.  & $B_{\rm T}^{0\,(e)}$ & $f_{60}$ & $f_{100}$& Notes$^{(f)}$\\
                & [Mpc]           &[$10^{-12}$ erg/s] &[$10^{-12}$ erg/s] &         &                      &  [Jy]    &  [Jy]    &           \\
\noalign{\smallskip}
\hline
\noalign{\smallskip}
NGC~0055$^{\rm R}$    &  1.3$^{[26]}$ & 1.8$^{[2,20]}$&  0.68$^{[2,20]}$  &ROSAT, ASCA& 7.63 &  77.00$^{[8]}$ & 174.09$^{[8]}$ &             \\
NGC~0253$^{\rm GR}$   &  3.0$^{[26]}$ & 2.5$^{[1,20]}$&  5.0$^{[1,20]}$   &  SAX    & 7.09   & 967.81$^{[8]}$  &1288.15$^{[8]}$ &            \\
NGC~0628              &  9.7$^{[26]}$ & ....          &  0.249$^{[15]}$   &  ASCA   & 9.76   &  21.54$^{[8]}$  &  54.45$^{[8]}$ &            \\
NGC~0891$^{\rm R}$    &  9.6$^{[26]}$ & 0.83$^{[20]}$ &  1.9$^{[20]}$     &  ASCA   & 9.37   &  66.46$^{[8]}$  & 172.23$^{[8]}$ &            \\
NGC~1569$^{\rm R}$    &  1.6$^{[26]}$ & 0.54$^{[3]}$  &  0.22$^{[3]}$     &  ASCA   & 9.42   &  54.36$^{[8]}$  &  55.29$^{[8]}$ &            \\
NGC~1808$^{\rm R}$    & 10.8$^{[26]}$ & 0.65$^{[20]}$ &  0.76$^{[20]}$    &  ASCA   &10.43   & 105.55$^{[8]}$  & 141.76$^{[8]}$ &            \\
NGC~2146$^{\rm R}$    & 17.2$^{[26]}$ & 0.82$^{[5]}$  &  1.11$^{[5]}$     &  ASCA   &10.58   & 146.69$^{[8]}$  & 194.05$^{[8]}$ &            \\
NGC~2276$^{\rm R}$    & 36.8$^{[26]}$ & 0.21$^{[20]}$ &  0.44$^{[20]}$    &  ASCA   &11.75   &  14.29$^{[8]}$  &  28.97$^{[8]}$ &            \\
NGC~2403$^{\rm R}$    &  4.2$^{[26]}$ & 1.6$^{[20]}$  &  0.93$^{[20]}$    &  ASCA   & 8.43   &  41.47$^{[8]}$  &  99.13$^{[8]}$ &            \\
NGC~2782              & 37.3$^{[26]}$ & 1.3$^{[25]}$  &  ....             &  ROSAT  &12.01   &   9.17$^{[8]}$  &  13.76$^{[8]}$ &            \\
NGC~2903$^{\rm R}$    &  6.3$^{[26]}$ & 0.79$^{[20]}$ &  0.686$^{[15]}$   &  ASCA   & 9.11   &  60.54$^{[8]}$  & 130.43$^{[8]}$ &            \\
NGC~3034$^{\rm GR}$   &  5.2$^{[26]}$ & 5.8$^{[21]}$  & 15.5$^{[21]}$     &  RXTE   & 5.58   &1480.42$^{[8]}$  &1373.69$^{[8]}$ &     = M~82 \\
NGC~3079              & 20.4$^{[26]}$ & 4.58$^{[2]+}$  &  0.78$^{[2]}$     &ROSAT, ASCA&10.41 &  50.67$^{[8]}$ & 104.69$^{[8]}$ &            \\
NGC~3256$^{\rm GR}$   & 37.4$^{[26]}$ & 0.69$^{[16]}$ &  0.586$^{[16]}$   &  ASCA   & 8.34   & 102.63$^{[8]}$  & 114.31$^{[8]}$ &            \\
NGC~3310              & 18.7$^{[26]}$ & 0.74$^{[20]}$ &  0.21$^{[27]}$    &  ASCA   &10.95   &  34.56$^{[8]}$  &  48.19$^{[8]}$ &            \\
NGC~3367$^{\rm R}$    & 43.6$^{[26]}$ & 0.18$^{[20]}$ &  0.16$^{[20]}$    &  ASCA   &11.92   &   6.44$^{[8]}$  &  13.48$^{[8]}$ &            \\
NGC~3556$^{\rm R}$    & 14.1$^{[26]}$ & 0.44$^{[20]}$ &  0.60$^{[20]}$    &  ASCA   & 9.83   &  32.55$^{[8]}$  &  76.90$^{[8]}$ &    = M~108 \\
NGC~3628              &  7.7$^{[26]}$ & 4.16$^{[2]+}$  &  0.98$^{[2]}$     &ROSAT, ASCA & 9.31&  54.80$^{[8]}$& 105.76$^{[8]}$ &             \\
Arp~299$^{\rm GR}$    & 41.6$^{[17]}$ & 0.57$^{[20]}$ &  1.08$^{[20,23]}$ &  ASCA   &11.85   & 113.05$^{[8]}$  & 111.42$^{[8]}$ &= NGC~3690 + IC~694\\
                      &               &               &                   &         &        &                 &                &{\small AGN at $\magcir$10 keV}$^{[32]}$\\
NGC~4038/39$^{\rm GR}$& 25.4$^{[26]}$ &0.72$^{[20,23]}$& 0.53$^{[20,23]}$  &  ASCA  &10.62   &  45.16$^{[8]}$  &  87.09$^{[8]}$ & = The Antennae\\
NGC~4449$^{\rm R}$    &  3.0$^{[26]}$ & 0.826$^{[4]}$ &  0.482$^{[4]}$    &  ASCA   & 9.94   &  36.0$^{[11]}$  &  73.0$^{[11]}$ &            \\
NGC~4631$^{\rm R}$    &  6.9$^{[26]}$ & 40.0$^{[2]+}$  &  1.15$^{[2]}$     &ROSAT, ASCA& 8.61 &  85.40$^{[8]}$  & 160.08$^{[8]}$ &           \\
NGC~4654$^{\rm R}$    & 16.8$^{[26]}$ & 0.06$^{[20]}$ &  0.09$^{[20]}$    &  ASCA   &10.75   &  13.39$^{[8]}$  &  37.77$^{[8]}$ &            \\
NGC~4666              & 14.1$^{[26]}$ & 0.16$^{[19]}$ &  0.29$^{[19]}$    &  SAX    &10.68   &  37.11$^{[8]}$  &  85.95$^{[8]}$ &            \\
NGC~4945$^{\rm G}$    &  5.2$^{[26]}$ & 1.3$^{[10]}$  &  5.4$^{[10]}$     &  SAX    & 7.43   & 625.46$^{[8]}$  &1329.70$^{[8]}$ & {\small AGN} \\
                      &               &               &                   &         &        &                 &                & {\small at $\magcir$10 keV}$^{[10,31,32,22,24]}$  \\
NGC~5236$^{\rm G}$    &  4.7$^{[26]}$ & 3.5$^{[18]}$  &  4.7$^{[18]}$     & ASCA    & 7.98   & 265.84$^{[8]}$  & 524.09$^{[8]}$ &  = M~83    \\
NGC~5253$^{\rm G}$    &  3.2$^{[26]}$ & 0.32$^{[14]}$ &  ....             &ROSAT    &10.47   &  29.84$^{[8]}$  &  30.08$^{[8]}$ &            \\
NGC~5457$^{\rm R}$    &  5.4$^{[26]}$ & 0.54$^{[20]}$ &  0.68$^{[20]}$    & ASCA    & 8.21   &  88.04$^{[8]}$  & 252.84$^{[8]}$ &  = M~101   \\
NGC~6946$^{\rm R}$    &  5.5$^{[26]}$ & 3.0$^{[20]}$  &  1.2$^{[20]}$     & ASCA    & 7.78   & 129.78$^{[8]}$  & 290.69$^{[8]}$ &            \\
NGC~7469$^{\rm G }$   & 65.2$^{[17]}$ & ....          & 29.8$^{[9]}$      & ASCA    &12.64   &  27.33$^{[8]}$  &  35.16$^{[8]}$ & AGN$^{[28,29,30]}$   \\
NGC~7552$^{\rm G }$   & 19.5$^{[26]}$ & 1.0$^{[13]}$  &  0.36$^{[13]}$    &Einstein &11.13   &  77.37$^{[8]}$  & 102.92$^{[8]}$ &            \\
NGC~7679              & 71.0$^{[6]}$ & 3.3$^{[6]}$    &  6.0$^{[6]}$      & SAX     &12.89   &   7.40$^{[8]}$  &  10.71$^{[8]}$ & AGN$^{[6]}$\\
IC~342$^{\rm R}$      &  3.9$^{[26]}$ & 1.8$^{[20]}$  &  11.0$^{[20]}$    &ASCA     & 6.04   & 180.80$^{[8]}$  & 391.66$^{[8]}$ &            \\

\noalign{\smallskip}
\hline
\hline
\end{tabular}
\end{flushleft}
\smallskip

$^{(a)}$ References: [1] Cappi et al. 1999; [2] Dahlem et al. 1998; [3] Della Ceca et al. 1996; [4]  Della Ceca et al. 
1997; [5] Della Ceca et al. 1999; [6] Della Ceca et al. 2001; [7] de Naray et al. 2000; [8] Sanders et al. 2003; [9] 
Guainazzi et al. 1994; [10] Guainazzi et al. 2000; [11] Hunter et al. 1986; [12] Iwasawa et al. 1993; [13] Maccacaro 
\& Perola 1981; [14] Martin \& Kennicutt 1995; [15] Mizuno et al. 1998; [16] Moran et al. 1999; [17] NED; [18] Okada
et al. 1990; [19] Persic et al. 2003; [20] Ranalli et al. 2003; [21] Rephaeli \& Gruber 2002; [22] Done et al. 2003; 
[23] Sansom et al. 1996; [24] Schurch et al. 2002; [25] Schulz et al. 1998; [26] Tully 1988; [27] Zezas et al. 1998;
[28] Perez-Olea \& Colina 1996; [29] Nandra et al. 2000; [30] Blustin et al. 2003; [31] Madejski et al. 2000; [32] 
Della Ceca et al. 2002.

$^{(b)}$ The superscripts G, R indicate whether an object is included in the Genzel 
et al. (1998) or Ranalli et al. (2003) samples, respectively.

$^{(c)}$ Distances are taken from Tully (1988) if $cz \leq 3000$ km s$^{-1}$, or are 
computed consistently assuming H$_0=75$ km s$^{-1}$ Mpc$^{-1}$ otherwise.

$^{(d)}$ Fluxes marked with a cross refer to the 0.1-2.0 keV band.

$^{(e)}$ Blue apparent magnitudes, corrected to face-on and for Galactic absorption, 
from RC3 (de Vaucouleurs et al. 1991).

$^{(f)}$ Other name(s); spectrally dominating component(s) at energies $\magcir$10 keV.
\end{table*}

\begin{table*}
\caption[] { 
Data III: 
The sample of {\it Hubble} Deep Field North galaxies (HDFN sample)$^{(a)}$.}
\begin{flushleft}
\begin{tabular}{ l  l  l  l  l  l  l  l }
\noalign{\smallskip}
\hline
\hline
\noalign{\smallskip}
Source &  $z$  &$F_{1.4\, {\rm GHz}}$& $L_{1.4\, {\rm GHz}}$ &      SFR$^{(b)}$    & $f_{2-10}$             &  $L_{2-10}$            &  Instr. \\
       &       &       [$\mu$Jy]     &[erg s$^{-1}$Hz$^{-1}$]&[$M_\odot$ yr$^{-1}$]&[erg s$^{-1}$cm$^{-2}$] &  [erg s$^{-1}$]        &         \\
\noalign{\smallskip}
\hline
\noalign{\smallskip}
134    & 0.456 &         210         & $1.03 \times 10^{30}$ &      25.8           &$~~2.8 \times 10^{-16}$ & $ 1.38 \times 10^{41}$ & Chandra \\
136    & 1.219 &         180         & $8.68 \times 10^{31}$ &     217.1           &$~~1.9 \times 10^{-16}$ & $ 9.17 \times 10^{41}$ & Chandra \\
148    & 0.078 &          96         & $1.16 \times 10^{28}$ &       0.3           & $<4.1 \times 10^{-17}$ & $<4.95 \times 10^{38}$ & Chandra \\
188    & 0.410 &          83         & $3.23 \times 10^{29}$ &       8.1           &$~~5.8 \times 10^{-17}$ & $ 2.26 \times 10^{40}$ & Chandra \\
194    & 1.275 &          60         & $3.24 \times 10^{30}$ &      80.9           &$~~2.0 \times 10^{-16}$ & $ 1.08 \times 10^{42}$ & Chandra \\
246    & 0.423 &          36         & $1.50 \times 10^{29}$ &       3.8           &$~~7.5 \times 10^{-17}$ & $ 3.13 \times 10^{40}$ & Chandra \\
278    & 0.232 &         160         & $1.84 \times 10^{29}$ &       4.6           &$~~1.6 \times 10^{-16}$ & $ 1.84 \times 10^{40}$ & Chandra \\

\noalign{\smallskip}
\hline
\hline
\end{tabular}
\end{flushleft}
\smallskip

$^{(a)}$ All data are taken, or derived, from Ranalli et al. (2003). Distant dependent 
quantities have been computed assuming H$_0=75$ km s$^{-1}$ and $q_0=0.5$. 

$^{(b)}$ The SFR has been computed using Condon's (1992) formula ${\rm SFR}(>5\, 
M_\odot$) = $L_{1.4\, {\rm GHz}} / (4 \times 10^{28} {\rm erg~s^{-1} Hz^{-1}} ) \, 
M_\odot {\rm yr}^{-1}$, which yields the SFR for stars in the mass range 
$5 \leq M/M_\odot \leq 100$ (assuming a Salpeter-like stellar IMF with 
$dN/dM \propto M^{-2.5}$, and a non-thermal radio spectral index $\alpha=0.8$).
\end{table*}

\section{Tracers of ongoing SFR} 

As mentioned above, the indicators of current SFR are stellar objects or systems whose 
activity lifetimes are very short compared to characteristic SF timescales. Based on their 
spectra, luminosity, and number, HMXBs are most appropriate SFR gaugers in the 2-10 keV 
band. In these stellar systems the donor is a massive ($M \magcir 8\, M_\odot$) OB star. 
Now, as a population OB stars are not important contributors to the X-ray output of galaxies 
(e.g., PR02), but their copious UV emission heats up dusty molecular clouds, with the heat 
emitted in the FIR. Thus, the 2-10 keV emission of HMXBs and the FIR emission of warm dust 
clouds are both related to the presence of short-lived OB stars (either isolated or in 
interacting pairs), with the emission ceasing soon after their death. In principle, then, 
the clouds' FIR luminosity and the 2-10 keV luminosity of HMXBs can be used independently to 
measure the instantaneous SFR. The crucial step is the separation of the cloud FIR luminosity 
and the HMXB X-ray luminosity from the corresponding total luminosities: if this can be done 
properly, the FIR-derived and the X-derived estimates of the SFR should agree. 

Massive OB stars found in HMXBs usually end up as supernovae. Shock waves triggered by 
SN explosions efficiently accelerate electrons to relativistic energies (e.g., 
Blandford \& Ostriker 1980). Electron energy losses are radiative synchrotron-Compton 
at energies higher than $\sim 300$ MeV, with the relative significance of each of 
these processes depending on the value of the mean magnetic field and the scattering 
radiation field (either FIR or CMB; see, e.g., Rephaeli et al. 1991). Typical radiative 
energy loss times are $t_{\rm loss} \sim 10^8$ yr (e.g., Rephaeli 1979), i.e., 
comparable to or longer than a typical SB duration. Therefore, galactic radio 
(synchrotron) emission provides a measure of the SFR averaged over similarly long 
timescale, and has little to do with the ongoing SFR except in galaxies with 
constant SFR (see section 3). In the following we shall deal with 2-10 keV luminosity 
of HMXBs and the FIR luminosity of dust clouds, both of which are {\it immediate} 
outcomes (and hence tracers) of the current SF activity.

\subsection{Infrared} 

In normal disk galaxies, the relationship between the FIR luminosity and the 
SFR is complex because stars with a variety of ages can contribute to dust 
heating, and only a fraction of the bolometric luminosity of the young 
stellar population is absorbed by dust (e.g., Lonsdale Persson \& Helou 1987; 
Walterbos \& Greenawalt 1996). 
In SFGs studied here, the physical coupling between the SFR and the IR 
luminosity is much more direct. Young stars dominate the radiation field that 
heats the dust, and the dust optical depths are so large that almost all of 
the bolometric luminosity of the SB is reradiated in the infrared. This makes 
it possible to derive a reasonable quantitative measure of the SFR from the 
FIR luminosity. 

A popular calibration of the SFR--$L_{\rm FIR}$ conversion (Kennicutt 1998) 
is based on the SB synthesis models of Leitherer \& Heckman (1995), which 
trace the temporal evolution of the bolometric luminosity for a given SFR, 
metal abundance, and IMF. Kennicutt (1998) computed the SFR calibration using 
their "continuous SF" models, in which the SFR is presumed to remain constant 
over the lifetime of the burst. The models show that the $L_{\rm bol}$/SFR ratio 
evolves relatively slowly between ages of 10 and 100 Myr, the relevant range 
for most of SBs (e.g., Bernl\"ohr 1993; Engelbracht 1997). Adopting the mean 
luminosity for the 10-100 Myr continuous bursts, solar abundances, and the 
Salpeter (1955) stellar initial mass function (IMF) defined as $dN/dM \propto 
M^{-2.35}$ in the mass range 0.1-100 $M_\odot$, and assuming the bolometric 
luminosity to be re-radiated by dust, yields 
   \footnote{ The FIR flux is defined (Helou et al. 1985) as a combination 
    of the {\it IRAS} $60\mu m$ and $100\mu m$ fluxes according to $f_{\rm 
    FIR} \equiv 1.26 \times 10^{-11} (2.58\, f_{60} + f_{100})$ erg s$^{-1}$ 
    cm$^{-1}$, where $f_{60}$ and $f_{100}$ are expressed in Jy.}
(Kennicutt 1998):
$$
{\rm SFR}(\geq 0.1 \, M_\odot) ~ = 
~ {L_{\rm FIR} \over 5.8 \times 10^9 L_\odot} ~ M_\odot {\rm yr}^{-1} \,.
\eqno(1)
$$
This lies within the range of published calibrations ($1-3 \times 10^{-10} M_\odot$ 
yr$^{-1}$ (e.g.: Lehnert \& Heckman 1996; Meurer et al. 1997; Devereux \& Young 1991; 
see also Inoue et al. 2000 for an analytical derivation, and Silva et al. 1998). 

Kennicutt's (1998) relation in eq.(1) is valid if all the FIR emission 
is due to warm molecular clouds heated up by newly born massive stars: i.e., eq.(1) 
properly applies to star-forming regions. If in addition to the warm component there 
is a "cirrus" component, i.e. emission from (colder) interstellar medium (ISM) heated by 
the general galactic UV radiation field which is powered also by relatively old stars, 
then a systematic bias will affect any SFR estimate based on eq.(1). Any cirrus 
component should then be filtered out from the observed FIR luminosity, in order to 
determine the warm emission arising from SF activity. This correction is especially 
important for objects that are not SB-dominated, such as local normal galaxies and 
SBGs (where the SB affects only the central disk region), for which the cirrus 
component is expected to be relatively important. In order to do so we follow David 
et al. (1992) in using the results of Devereux \& Eales (1989): Assuming that the O 
stars responsible for FIR emission are (ultimately) linked to particle acceleration 
(via SN shocks) and (in turn) to nonthermal radio emission, it would be predicted that 
the cirrus component should be directly proportional to $L_{\rm B}$ (with the latter 
a measure of the galactic stellar content), i.e. $L_{\rm FIR}^{\rm SB} = L_{\rm FIR} 
- x \,L_{\rm B}$. Devereux \& Eales (1989) find that if $x=0.14$, then for their sample 
of 237 galaxies the FIR--radio (i.e., $L_{\rm FIR}^{\rm SB}$--$L_{1.4\, {\rm GHz}}$) 
correlation is optimized and becomes linear. This implies that, on average, $14\%$ of 
the blue luminosity of a normal or starbursting galaxy is reradiated in the FIR band 
as cirrus emission: for this class of galaxies (see Table 2), the FIR luminosity to be 
used in eq.(1) is then $L_{\rm FIR}^{\rm SB}$ (which we here still refer to as $L_{\rm 
FIR}$ for simplicity). Anticipating results obtained later in this paper, we can state 
that allowing for cirrus emission, while theoretically required and practically favorable 
(doing so only strengthens our results), amounts only to a minor correction.

\subsection{X-Rays}

Being able to ascertain whether the 2-10 keV band is dominated by HMXBs or LMXBs 
may yield substantial insight on the SF history of a galaxy. In a SB environment 
the formation of low-mass stars could be suppressed if SN blast waves of more 
rapidly forming massive stars disrupt the slowly forming less massive stars before 
these complete (or even reach) their Hayashi tracks. The ensuing stellar IMF would 
be top-heavy (see, e.g., Doane \& Mathews 1993 and Rieke et al. 1993 for M~82), in 
which case no LMXBs would form, and the resulting binary population of the SB 
consists only of HMXBs. But even if the IMF were not top-heavy, one isolated SB 
episode would not trigger an increase of the LMXB population. In fact, the time 
required for the $\mincir$1$\,M_\odot$ optical companion in a LMXB system to evolve 
out of the main sequence and come into Roche-lobe contact (and hence start the X-ray 
bright phase) substantially exceeds a typical SB lifetime ($\tau_{\rm SB} \mincir 
10^8$ yr). So, whatever is the mass range of the relevant stellar IMF, during one 
isolated SB episode there is time only for HMXBs (i.e., by accretion-powered X-ray 
pulsars) to form (in larger numbers for a top-heavy IMF). The resulting 2-10 keV 
synthetic spectrum would then be dominated by HMXBs, and hence the SB spectral slope 
would be characterized by a flat photon index $\Gamma \simeq 1.2$ (see Fig.1-{\it 
right}).
	\footnote{ Spectral components that are flat in the 2-10 keV band 
	can come from AGNs. Both a direct nuclear emission (with $\G \sim 2$) 
	that is heavily absorbed, or an emission that is reflected by a warm 
	medium, are observed as relatively flat ($\G \sim 1.2$) spectra (e.g., 
	NCC~6240, NGC~4945, Arp 299: Vignati et al. 1999, Guainazzi et al. 
	2000, Della Ceca et al. 2002). In general there will be no degeneracy 
	between the AGN and HMXB interpretations of the flat spectral components, 
	as the inferred luminosity should enable a clear identification. }

In the case of recurrent SF bursts, LMXBs could be important contributors in the 2-10 keV 
band. This may happen in, e.g., galaxies that are members of pairs with highly eccentric 
orbits, or that are found in crowded environments (e.g., compact groups: see Hickson et al. 
1989; see also Della Ceca et al. 1997). In both cases, tidal interactions would be recurrent, 
so bright LMXB could be present as leftovers from earlier interactions. If one such galaxy is 
observed in a post-SB (or quiescent) phase, the 2-10 keV emission would practically come only 
from LMXBs. 

Therefore, determining the type of X-ray binaries whose emission dominates the 2-10 keV 
luminosity of SFGs could provide a clue both to the understanding of SF history of those 
galaxies (e.g., Holt et a. 2003), and to using $L_{2-10}$ as an indicator of the ongoing 
SFR. As an example of the latter, let us consider three representative cases of SFGs. 

\noindent
{\it (i)} In a galaxy where the SF activity has been constant over several $10^8$ yr, 
the relative frequencies of occurrence of stellar X-ray sources (SNRs, HMXBs, LMXBs) 
are expected to agree with those predicted for the Galaxy (which has a SFR of $\sim 2\, 
M_\odot$ yr$^{-1}$) by the Iben et al. (1995a,b) models, upon which the PR02 template 
X-ray spectrum is based. Comparing the deduced HMXB content of a given galaxy (from 
X-ray spectral decomposition) with that of the Galaxy, its current SFR can be estimated. 

\noindent
{\it (ii)} For SB-dominated galaxies the above considerations suggest that $L_{2-10}$ is 
an indicator of SFR. 

\noindent
{\it (iii)} In the intermediate case of a galaxy undergoing a minor SB (which is usually 
located in the central region), $L_{2-10}$ is a superposition of the emissions from the 
(quiescent) bulge/disk and the ongoing SB. Only after these two components are separated 
out, can the 'quiescent' and the 'bursting' SFRs be estimated (see Fig.2).

\section{The samples} 

To test the idea that the SFR-$L_{\rm x}$ relation largely reflects the ability to use  
HMXBs to trace the current SFR, we have selected three representative samples of SFGs. 

The sample of ULIRGs (Table 1) -- sources with bolometric luminosity $>10^{12}L_\odot$ 
emitted mostly in the IR (8-1000$\mu m$) band -- is essentially the list of IRAS-selected 
galaxies that represent some of the most luminous galaxies in the local Universe observed 
with ISO by Genzel et al. (1998), whose FIR selection makes it unbiased with respect to 
absorption. It is flux-limited and complete down to $S_{60 \mu m} \geq 5.4$ Jy, and 
high-quality IR/optical spectroscopic data are available for all the sources (Lutz et al. 
1999; Veilleux et al. 1999). X-ray data in both the soft and hard bands are from {\it ASCA}, 
{\it BeppoSAX}, and {\it XMM-Newton}. According to the most extensive spectral survey of 
ULIRGs to date (Franceschini et al. 2003), the ULIRG spectra are qualitatively similar to 
SBG spectra, plus possibly (in $40\%$ cases) a heavily absorbed PL component that shows up 
at $>$5 keV (rest-frame). A similar incidence rate of AGNs in ULIRGs is deduced using 
mid-IR diagnostics (Genzel et al. 1998).

The sample of SBGs (Table 2), compiled from Genzel et al. (1998) and Ranalli et al. (2003), 
consists of local FIR-bright objects known to contain central starbursts of various strengths 
for which X-ray spectral information is available (from {\it Einstein}, {\it ROSAT}, {\it 
ASCA}, {\it BeppoSAX}, and {\it RossiXTE}). It should be stressed that this sample comprises 
a fair number of 'normal' galaxies, i.e. galaxies whose SFR is comparable to that of the Milky 
Way. These SFGs are useful for checking the SFR--$L_{\rm x}$ relation at the low-SFR end of 
the galaxy distribution. A more appropriate name for this sample should then be 'local SFGs' 
(of which SBGs are, technically speaking, a subset); but for lack of a definite, clear-cut 
distinction between 'normal' and 'starburst' galaxies, in this paper we'll stick to the 
denomination 'SBG sample' in order to emphasize that our main interest for these galaxies lies 
in their SF activity. Judging by currently available X-ray measurements, most of these galaxies 
share a remarkable spectral homogeneity with their spectra usually described (most simply) as a 
soft thermal component at $<1$ keV, plus a hard (sometimes cutoff) PL at $>2$ keV (e.g., Dahlem 
et al. 1998). 

The sample of {\it Hubble} Deep Field North (HDFN) galaxies (Table 3) was 
selected by Ranalli et al. (2003) from the 1 Ms {\it Chandra} (Brandt et al. 2001) 
and radio (1.4 GHz: Richards 2000 and Garrett et al. 2000) catalogues of the HDFN, 
which reaches limiting fluxes low enough for SFGs at $z \sim 1.3$ to be detected. 
As emphasized by Ranalli et al. (see also Grimm et al. 2003), these galaxies can be 
used to investigate the high-$z$ behaviour of the SFR--X-ray luminosity relation, 
which is being explored primarily in the local universe ($z \mincir 0.1$). Lacking 
FIR data, we estimate the SFR from the radio flux density at 1.4 GHz using the 
relation ${\rm SFR}(\geq 5\, M_\odot) =L_{1.4\,{\rm GHz}}/ (4 \times 10^{28} \, 
{\rm erg~s^{-1} Hz^{-1}}) M_\odot {\rm yr}^{-1}$ (Condon 1992), with the resulting 
SFR are reported in Table 3. It should be emphasized that the Condon formula yields 
the formation rate of stars with $5 \leq M/M_\odot \leq 100$. To compare these SFR 
estimates with those we use for the SBG and ULIRG samples, we must extrapolate the 
radio SFR down to $0.1\,M_\odot$ using the Salpeter (1955) stellar IMF upon which 
the Kennicutt (1998) formula is based. Doing so the radio and FIR estimates will be 
consistent
	\footnote{In the derivation of the SFR--$L_{1.4, {\rm GHz}}$ conversion, 
	Condon (1992) used a Salpeter-like IMF with index 2.50, instead of 2.35 
	as adopted by Kennicutt (1998). For our purposes here this difference 
	is negligible.}. 
We then obtain ${\rm SFR}(\geq 0.1\, M_\odot) = 5.5 \,{\rm SFR}(\geq 5\, M_\odot)$. {\it 
We will use these new values}, ${\rm SFR}(\geq 0.1\, M_\odot)$, {\it as the reference SFR 
values for our HDFN galaxies.} It should be pointed out that, owing to the long lifetime 
of the radio-emitting relativistic electrons, $\sim$10$^9$ yr, radio emission may not be able 
to trace shorter-timescale SBs (see section 2). However, it is realistic to assume that HDFN 
galaxies, being in an earlier evolutionary phase than local galaxies, are dominated by strong 
continuous SF activity (e.g.: Madau et al. 1996; Thompson et al. 2001) and hence their SFR 
can be effectively measured by the non-thermal radio emission.

\begin{figure}
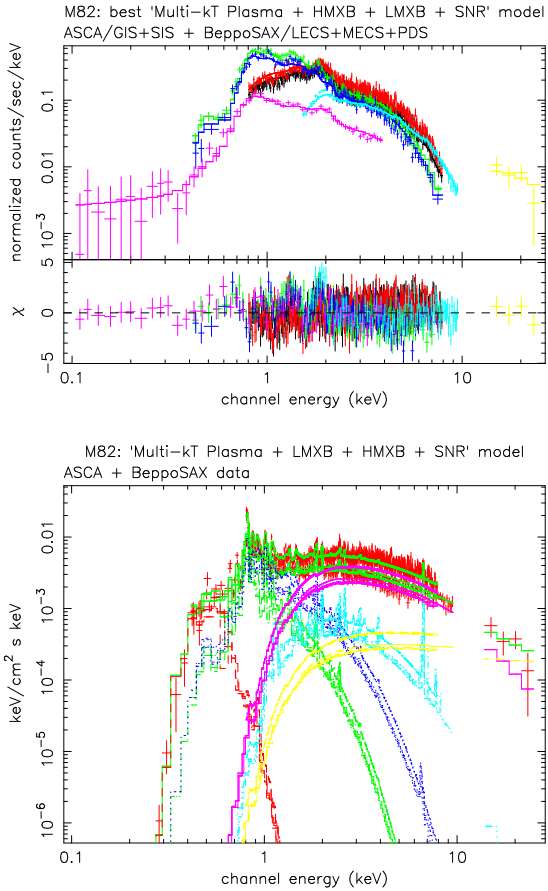


\vspace{3.4cm}
\includegraphics{0500fig2a.cps}
\vspace{.5cm} 
\vspace{8.0cm}
\includegraphics{0500fig2b.cps}
\caption{ The "standard" model shown in Fig.1-{\it left} is fitted to the combined {\it 
ASCA} and {\it BeppoSAX} data of M~82 ({\it ASCA}: black, red, green, and dark-blue points; 
{\it BeppoSAX}: purple, cyan, and yellow points). The profiles of the spectral components 
are kept frozen in the fitting procedure (while their amplitudes are left free to vary) 
and ({\it bottom}) correspond to: photon index $\G=1.2$ for HMXBs (yellow); photon index 
$\G=1.4$ and cutoff energy $\epsilon_{\rm c} = 7.5$ keV for LMXBs (purple); temperature 
$kT=2$ keV and chemical abundance $Z=Z_\odot$ for SNRs (cyan). At lower energies, three 
plasma components (having $kT=0.065$, 0.45, 0.75 keV; and $Z=0.1\, Z_\odot$) are required 
(red, green, and dark blue). Shown are the model fitted to the data ({\it top}, energy 
spectrum), the residuals of the fit ({\it middle}), and the unfolded energy spectrum 
({\it bottom}). }

\end{figure}

\begin{figure}
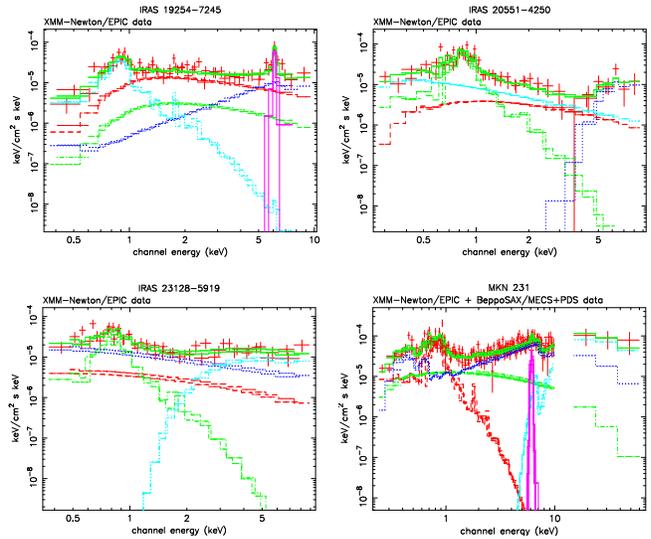


\vspace{4.0cm}
\includegraphics{0500fig3a.ps}
\hspace{1.3cm}
\includegraphics{0500fig3b.ps}
\vspace{.5cm} 
\vspace{2.6cm}
\includegraphics{0500fig3c.ps}
\hspace{1.3cm}
\includegraphics{0500fig3d.ps}
\caption{ 
The energy spectra of the four AGN-contaminated ULIRGs of the Franceschini et al. (2003) 
sample. The {\it XMM}-Newton data for IRAS~19254-7245, IRAS~20551-4250 and IRAS~23128-5919 
span the $\sim$0.4-10 keV energy range, while the composite {\it XMM}-Newton + {\it BeppoSAX} 
data for Mkn~231 span the 0.2-60 keV energy range. The HMXB contributions have been estimated 
by perturbing the published best-fit spectral models with an additional $\Gamma=1.2$ PL. Such 
additional components are reported in the four panels (clockwise from top left) as green, red, 
green, and red lines. 
}

\end{figure}

\section{HMXBs as gauges of the SFR} 

We now demonstrate that the HMXB portion of the 2-10 keV luminosity, $L_{2-10}^{\rm HMXB}$, 
can be used as a gauge of the SFR. To this end we compare the SFR computed from the FIR according 
to the Kennicutt (1988) formula [see eq.(1)] with the SFR estimated from the integrated luminosity 
of the bright-HMXB population. We assume a mean luminosity of $5 \times 10^{37}$ erg s$^{-1}$ for 
bright HMXBs. For each sample galaxy, we estimate the number of HMXBs from the $L_{2-10}^{\rm HMXB}$. 
By comparison with the Galaxy, which hosts $\sim 50$ bright HMXBs (Iben et al. 1995a, and references 
therein), and has SFR $\sim 2 \,M_\odot$ yr$^{-1}$ (e.g., Matteucci 2002), we estimate the current 
SFR of the galaxy in question. This simple procedure is applied to both the ULIRG and SBG samples.
\medskip

\noindent
$\bullet$ {\it ULIRGs.} If $L_{2-10}$ is used, the X-ray estimates are in overall agreement with the 
FIR estimates for the AGN-free ULIRGs (Fig.4: filled dots), while they are clearly in excess for 
objects whose 2-10 keV emission is contaminated, or dominated, by a central AGN         
	\footnote{With both SFR estimates linear in the respective 
	luminosities, an equivalent statement is that, for a given 
	$L_{\rm FIR}$, AGN-contaminated ULIRGs are X-ray overluminous 
	compared with SB-dominated ones of the same $L_{\rm FIR}$; see also
	Ptak et al. 2003.}
(Fig.4: empty dots and crosses, respectively). The overall agreement between FIR-derived and 
X-derived estimates of the SFR suggests that the 2-10 keV luminosity of AGN-free ULIRGs is indeed 
(mainly) due to HMXB: this conclusion is in accord with their flat observed 2-10 keV continua, 
which are consistent with the $\G \sim 1.2$ spectra observed in Galactic HMXBs. 

For the objects in which the SB and AGN contributions to the 2-10 keV flux are comparable (see 
Table 1), we employed the following crude procedure in order to estimate the HMXB contribution, 
a procedure that is admittedly quite arbitrary given the uncertain observational situation, but 
that nonetheless has some theoretical justification. For the 4 objects taken from Franceschini 
et al. (2003), we have added a $\G \sim 1.2$ PL component to the published spectral models 
	\footnote{ IRAS~19254-7245: Braito et al. 2003; IRAS~20551-4250, 
	IRAS~23128-5919: Franceschini et al. 2003; and Mkn~231: Braito 
	et al. 2004.}, 
repeating the $\chi^2$ minimization. Results of the fitting, while not compelling because the 
quality of the data allows only simple model testing, do indicate a roughly $\sim 10\%$ HMXBs 
contribution to the 2-10 keV flux. Specifically, the additional PL fluxes are $f_{2-10} = 2.06$, 
1.89, 1.44, and $9.25 \times 10^{-14}$ erg cm$^{-2}$ s$^{-1}$, respectively (in the order listed 
in footnote 6), which in turn correspond to luminosities of $L_{2-10} = 1.49 \times 10^{41}$, 
$6.59 \times 10^{40}$, $5.44 \times 10^{40}$, and $3.15 \times 10^{41}$ erg s$^{-1}$ (see Fig.3). 
For Mkn~273 we have followed the argument of Xia et al. (2002) who, in their analysis of the {\it 
Chandra} spectrum, have identified a moderately absorbed PL component which they suggest must be 
at least partially associated with HMXBs. (This PL is in addition to a steeper and more absorbed
PL that Xia et al. associate with the the direct light from the central AGN.) Therefore, lacking 
more detailed information, we assume the HMXB contribution in Mkn~273 to be half of the flux in 
the less-absorbed PL, i.e. $f_{2-10} = 4.6 \times 10^{-14}$ erg cm$^{-2}$ s$^{-1}$ corresponding 
to $L_{2-10} = 1.27 \times 10^{41}$ erg s$^{-1}$. We emphasize that for the five AGN-contaminated 
ULIRGs of our sample the estimated HMXB contributions are very preliminary, given the substantial 
uncertainties. However, it is remarkable that the 'corrected' luminosities cluster so closely with 
the total 2-10 keV luminosities of the AGN-free objects (Fig.5-{\it right}: empty dots). 
\medskip

\noindent
$\bullet$ {\it SBGs}. In comparing SBGs and ULIRGs, we should bear in mind a main 
difference between the two classes that is particularly relevant to our discussion 
here. In ULIRGs the star forming activity is generally very intense and involves 
most of the stellar disks; in local SBGs (such as those included in Table 2), SF is 
considerably less intense and occurs mainly in the very central disk region. Thus, 
for non spatially resolved measurements, such as the {\it ASCA} and {\it BeppoSAX} 
data used in Figs.4 and 5, the 2-10 keV emission is expected to come mostly from 
HMXBs in ULIRGs, and from a mix of HMXBs (in the central starbursting region) and 
LMXBs (in the underlying quiescent disk) in normal and staburst galaxies. Therefore, 
if this effect is not corrected for, our X-ray-based estimate of the SFR, while 
fairly adequate in principle (albeit crude) in the case of ULIRGs, is systematically 
overestimated in the SBG sample by a factor $1/f$, with $f$ the HMXB-to-total 
luminosity ratio. Thus, we need to estimate $f$ for the SBG sample listed in Table 2. 

To this end we fit the template X-ray model of PR02 (see Fig.1-{\it left}), keeping the 
profiles of the spectral components frozen, and letting only the amplitudes free to vary, 
to available data for such galaxies. In principle this procedure should be applied to 
each and every galaxy in the sample. In practice, however, for most of the objects the 
available data are such that it is difficult even to determine whether the harder spectral 
component is PL or thermal (e.g., Dahlem et al. 1998; Dahlem et al. 2000), let alone a 
more complicated multi-component (thermal and PL) model including also emisssion from HMXBs, 
SNRs, and LMXBs. The very nearby M82, which has been observed with {\it ASCA} in the 0.4-10 
keV range (Ptak et al. 1997) and {\it BeppoSAX} in the 0.1-60 keV range (Cappi et al. 1999), 
is currently the best example of a relatively well-observed SBG. In Fig.2 we show the result 
of fitting the PR02 template to the combined {\it ASCA}+{\it BeppoSAX} data (see the 
Appendix). The results of these fits yield $f \sim 0.2$.

In order to check to what extent this value of $f$ may be taken as representative of SBGs in 
general, we have performed additional fits for a subset of our SBG objects (including M~83, 
NGC~253, NGC~2146, NGC~2903, NGC~3310, NGC~3256) being aware, however, that the archival 
{\it ASCA} spectra available for this analysis are inferior to that of M82, so any result 
from such a comparison is necessarily preliminary and merely suggestive. (For the other 
objects of the SBG sample for which {\it ASCA} data exist, we felt that the quality of the 
data was not sufficient for even a rough estimation of $f$.) In addition to one or more 
sub-keV thermal plasma components, the fitting models we used did include a HMXB component 
plus either a plain/cutoff PL (NGC~2146, NGC~2903, NGC~3256, NGC~3310), or the explicit 
SNR+LMXB components (M~83, NGC~253). Within the limits imposed by the relatively poor 
statistics of these data, the results are consistent with $f \sim 0.2$ for the objects with 
highest quality data (M~83, NGC~253, NGC~2903), while suggesting a slightly higher value of 
$f$ in the other cases (NGC~2146, NGC~3256). The exception seems to be NGC~3310, whose hard 
component is flat enough ($\Gamma \sim 1.4$: Zezas et al. 1998) to be accounted for by HMXBs 
alone, and hence $f \sim 1$. In conclusion, our additional spectral analysis supports the 
deduction that SBGs have $f \sim 0.2$, while showing a high scatter in the actual value of 
this factor.

The possibility that the emission of SBGs may be dominated by a few very luminous point sources 
(e.g., for M~82 see Kaaret et al. 2001 and Strohmayer \& Mushotzky 2003), and that analyses of 
spatially integrated SBG spectra may be biased, motivates further attempts to estimate $f$. 
Information about the relative emission from HMXBs can also be derived from studies on point 
source populations in nearby galaxies based on the high angular resolution {\it Chandra} data. 
Studies on individual local galaxies have yielded luminosity functions of X-ray point sources 
(XPLFs) down to limiting luminosities of $\sim$10$^{36-37}$ erg s$^{-1}$ (e.g., Kilgard et al. 
2002; Colbert et al. 2003; Fabbiano et al. 2001; Zezas et al. 2002, 2003; Roberts et al. 2002a; 
Soria \& Wu 2002; Bauer et al. 2001; Griffiths et al. 2000; Swartz et al. 2003; Kaaret 2002; 
Soria et al. 2003; Gao et al. 2003; Terashima \& Wilson 2003; Kong 2003). Such XPLFs can be 
described as $N(>L) \propto L^{-\alpha}$ with $\alpha \sim$ 0.5, 1, 1.5 for starburst, normal 
spiral, and elliptical galaxies (and bulges), respectively (Kilgard et al. 2002 and references 
therein). The difference in XPLFs among galaxy types, which implies that higher-SFR galaxies 
have relatively more sources (presumably, HMXBs) at high luminosities, is presumably related to 
the age of the underlying X-ray binary population. A simple birth-death model of X-ray binaries 
predicts a difference of 1 between the XPLF slope for an equilibrium XP population and that for 
an impulsively-formed quietly-ageing XP population. However, XP counts do not give clues as to 
the nature of the XPs. Prestwich et al. (2002) have proposed to classify XPs in external galaxies 
on the basis of their X-ray colors (defined using {\it Chandra}'s 0.65, 1.5, and 5 keV bands), 
finding highly significant differences in the X-ray colors of bulge and disk XPs, i.e. of LMXBs 
and of HMXBs and SNRs. Colbert et al. (2003) have analyzed 1441 XPs detected in 32 nearby 
galaxies, concluding that the XP luminosity is well correlated with the K-luminosity (i.e., the 
stellar mass) and the FIR+UV luminosity (i.e., the SFR), suggesting that XPs are connected to 
both the old and young stellar populations. Using their X-ray color diagrams, we find that 
$\sim$0.2 of the XPs observed in spiral and interacting galaxies can be classified as HMXBs. 
Correspondingly, a fraction $\sim$0.25 of the 0.3-8 keV XP luminosity of these galaxies can be 
attributed to such $\Gamma \sim 1.2$ sources. Because the XP luminosity typically accounts for 
$\sim 80\%- 90\%$ of the total luminosity (e.g., Roberts et al. 2002a), we deduce that (putative) 
HMXBs account for $\sim$20$\%$ of the integrated 0.3-8 keV luminosity. This value is in reasonable 
agreement with our spectral estimate of $f \sim 0.2$, mostly based on M~82.
	\footnote{The integrated XP luminosity of SFGs is dominated by ultra-luminous 
	X-ray sources (ULXs), i.e. off-nucleus objects with $L \geq 10^{39}$ erg s$^{-1}$ 
	which is the Eddington luminosity for a $8 \, M_\odot$ black hole (BH) (which is 
	the limiting BH mass obtainable via stellar evolution). The nature of ULXs is 
	still unclear. However, ULXs appear to be related mostly to high-SFR environments 
	(Fabbiano et al. 2001; Lira et al. 2002; Roberts et al. 2002a; Zezas et al. 2002, 
	2003; Humphrey et al. 2003; Gao et al. 2003; Karet et al. 2001; Strickland et al. 
	2001): the identified optical counterparts of ULXs are indeed O stars, so 
	suggesting a relation of ULXs with HMXBs (Liu et al. 2002; Roberts et al. 2002b). 
	The spectra of ULXs, reminiscent of those of Galactic BH X-ray binary candidates 
	in high state, are described by a model comprising an accretion disk with inner-edge 
	temperature of $\sim$1.1 keV plus a hard ($\Gamma \sim 1.2$) PL similar to that 
	observed in HMXBs (Zezas et al. 2002; Foschini et al. 2002; see also Fabbiano \& 
	White 2003). A scenario in which a population of HMXBs, with $\magcir$15$\,M_\odot$ 
	donors transferring mass on their thermal timescales (and hence at super-Eddington 
	rates) to $\magcir$10$\,M_\odot$ accretors, fits current observations on ULXs (King 
	2003). Our computation of $f \sim 0.2$, based on the XP colors corresponding to $\G 
	\sim 1.2$, does therefore incorporate the possible presence of ULXs.}
We therefore consider $f \sim 0.2$ to be the currently appropriate, representative {\it average} 
value, for the SBG sample.

An analogous correction to the FIR emission, aimed at removing any contribution that is unrelated 
to ongoing SF activity, is also in order. As discussed in section 2.1, we must subtract the 
(SB-unrelated) cirrus component from the FIR emission of SBGs 
	\footnote{
	Some objects in Table 2 are found at low Galactic latitudes ($|b| < 15$ 
	degrees). For these the corrections for foreground Galactic absorption 
	tend to be large and uncertain. An extreme case is IC 342 ($b=10.58$ deg), 
	for which the $B$-band absorption is estimated to be as large as 3.360 mag 
	(Burstein \& Heiles 1982) or 2.407 mag (Schlegel et al. 1998). As the 
	adopted statistical correction for cirrus emission turns out to be unphysical 
	in this case, we will leave IC342 out from further analyses involving 
	cirrus-corrected FIR luminosities. No result in this paper will depend on 
	whether this object is included in (or excluded from) the sample. [Note
	that IC~342 is uncharacteristically underluminous in FIR and soft X-rays 
	compared with hard X-rays, suggesting the presence of a HMXB population typical
	of a SBG, but with reduced FIR and hot gas emissions (Bauer et al. 2003)].
	}.
[The practical impact of this necessary correction is minor (see Fig.4), and thus has no 
bearing on any of our results.]

The SFR, as measured by $f L_{2-10}$, turns out to be in good agreement with that inferred 
from $L_{\rm FIR}$ (see Fig.5, filled squares
	\footnote{The SBG sample contains one object, Arp~299, which 
	was recently shown to host an AGN (Della Ceca et al. 2002). 
	Remarkably, this object lies in the region populated by the 
	AGN-dominated ULIRGs.});
the left panel includes the correction for the cirrus component in the FIR emission. To 
appreciate the effect of the correction, in the right panel we show the plot without the 
cirrus correction.
\medskip

\begin{figure}
\vspace{4.3cm}
\includegraphics{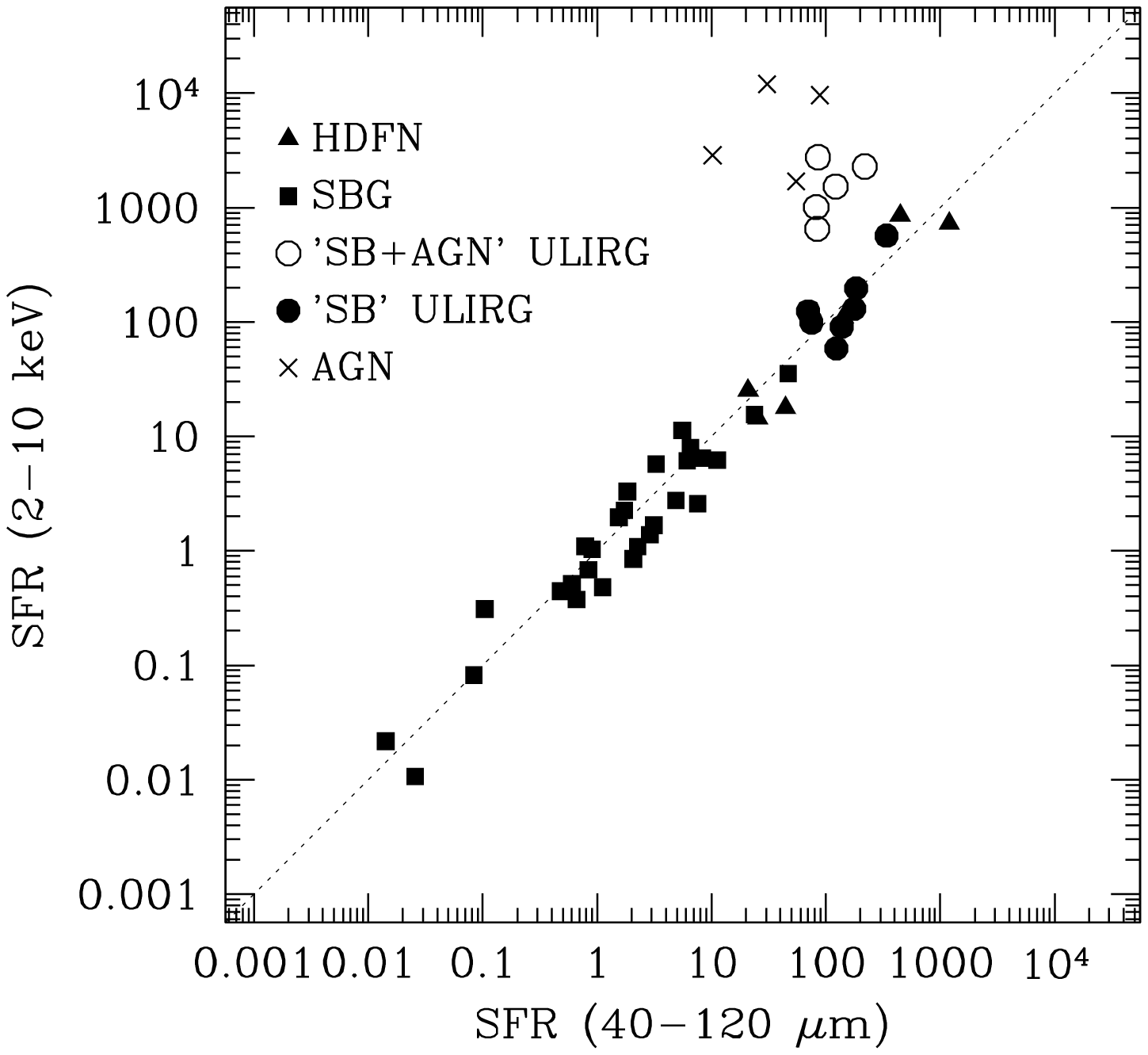}
\hspace{.5cm} 
\includegraphics{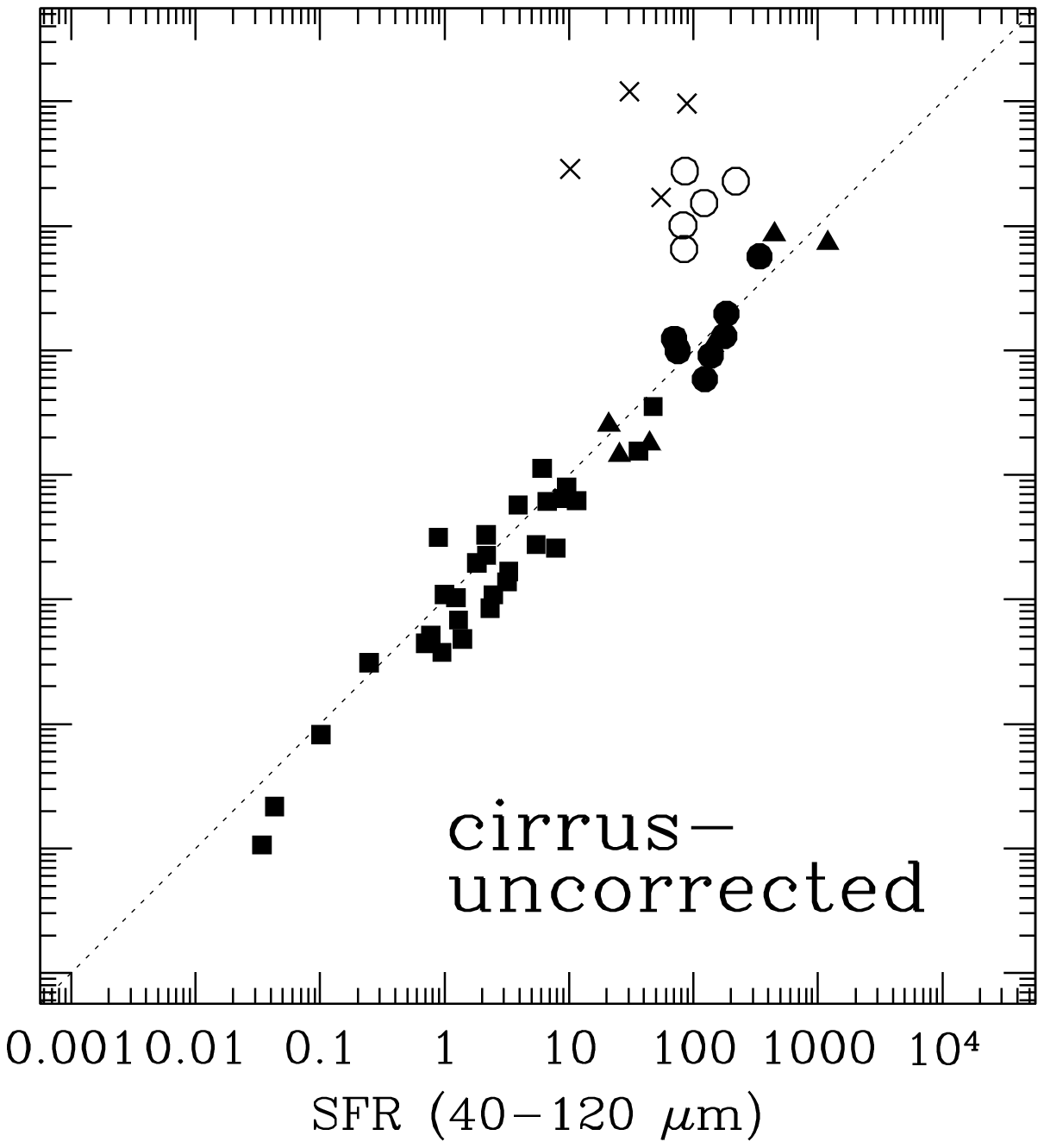}
\caption{
The SFR estimated from X-ray emission versus that estimated from FIR emission. For all 
objects it is assumed that the emission in both bands is related to starburst activity. 
Filled squares and filled dots represent, respectively, SBGs and SB-dominated ULIRGs; 
empty dots denote SB/AGN-powered ULIRGs, while crosses denote AGN-dominated objects; 
filled triangles denote HDFN galaxies (for which, due to the lack of FIR measurements, 
the SFRs have been estimated from radio fluxes). The dotted line, representing the 
SFR$_{\rm x}$ = SFR$_{\rm FIR}$ relation, is only meant to guide the eye. The right 
panel shows the case where the FIR luminosities of normal and starbursting galaxies 
are not corrected for cirrus emission.}
\end{figure}

\noindent
$\bullet$ {\it Hubble Deep Field North Galaxies}. In the SFR--$L_{2-10}$ plane the HDFN 
galaxies overlap (with a wider spread in SFR) with the SB-dominated ULIRGs (see Fig.5-{\it 
left})
	\footnote{Had we mistakenly used the plain SFR$(\geq 5\, M_\odot)$ 
	values resulting from applying Condon's (1992) formula to the radio 
	data  -- instead of the extrapolated SFR$(\geq 0.1\, M_\odot)$ 
	values --  the HDFN galaxies would have lined up, and partially 
	overlapped, with the SBGs in the SFR--$L_{2-10}$ plane.}. 

Lacking spectral information, a simple and realistic assumption is that HDFN galaxies -- 
likely `juveniles' -- are SF-dominated. Indeed Cohen (2003) finds that, based on the emitted 
luminosity in the 3727 {\AA} line of [O~II], the average SFR is about an order of magnitude 
higher in HDFN galaxies than in local ones. Furthermore, the age of a galaxy at $z \sim 1$, 
$\sim$6 Gyr, is not long enough for LMXBs -- which have sub-solar mass donors -- to form (see 
Maeder \& Meynet 1989). These considerations suggest that the 2-10 keV emission of HDFN 
galaxies is due essentially to HMXBs, like in the SB-dominated ULIRGs discussed above. We 
then set $L_{2-10}^{\rm HMXB} = L_{2-10}$. The X-ray estimates of the SFR agree comfortably 
with those inferred from the radio data (see Fig.4). Therefore, the SFR--$L_{2-10}^{\rm HMXB}$ 
relation for HDFN galaxies is consistent with that for SBGs and SB-dominated ULIRGs (see 
Fig.5-{\it right}). 
\medskip

Analysis of the three samples leads us to the following conclusions: 
\smallskip

\noindent
{\it i)} In 
SB-dominated ULIRGs the 2-10 keV luminosity is a 
gauge of the SFR.
This result lends further support to the interpretation of the $\sim$ 2-10 keV emission of 
ULIRGs as being (mainly) the integrated emission from HMXBs. In those cases where an AGN 
clearly shows up (Franceschini et al. 2003) when this component is subtracted out, the 
residual 2-10 keV luminosities imply SFR values that agree with the FIR-based ones.

\noindent
{\it i)} In 
local star-forming (i.e., normal and starburst) galaxies only a relatively minor fraction 
($f \sim 0.20$) of the 2-10 keV luminosity is related to the ongoing SF activity. Correcting 
for this bias results in essentially the same SFR-$L_{2-10}^{\rm HMXB}$ relation for both 
ULIRGs and SBGs.
      
\noindent
{\it iii)} Distant 
($z \sim 1$) SFGs have SFR values and 2-10 keV luminosities similar to those of ULIRGs. 
\medskip

Within the uncertainties of the data (most notably those involved in identifying the 
HMXB contribution in our sample of AGN-contaminated objects) the resulting relation is 
roughly linear, 
$$
{\rm SFR}(\geq 0.1 \, M_\odot) ~ = ~ {L_{2-10}^{\rm HMXB} \over 10^{39} {\rm erg~ s}^{-1} } 
~M_\odot {\rm yr}^{-1}
\eqno(2)
$$
with an estimated $20\%$ statistical error.
 
It should be remarked that the SFR-luminosity relation is not tight if no correction is 
applied (see Fig.5-{\it left}). Both SBGs and AGN-contaminated ULIRGs are overluminous for 
their SFR when compared with the SB-dominated ULIRGs and HDFN galaxies. Indeed, SBGs are 
contaminated by substantial LMXB emission, and in AGN-contaminated ULIRGs the AGN emission 
is typically brighter than the SB emission. After correcting for both types of excess, no 
hint of non-linearity appears in the SFR-luminosity relation (Fig.5-{\it right}). 

\begin{figure}
\vspace{4.3cm}
\includegraphics{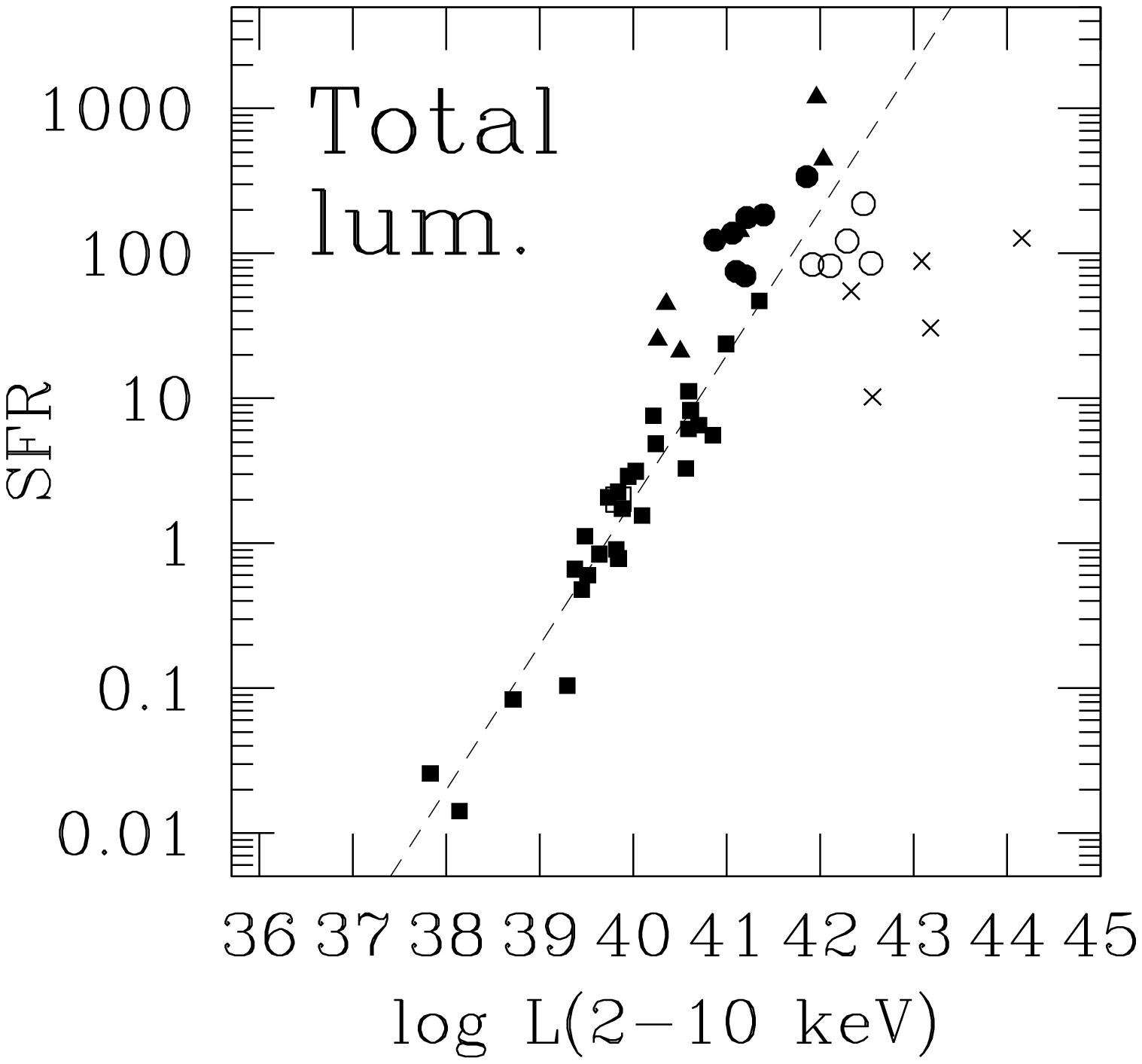}
\hspace{.5cm} 
\includegraphics{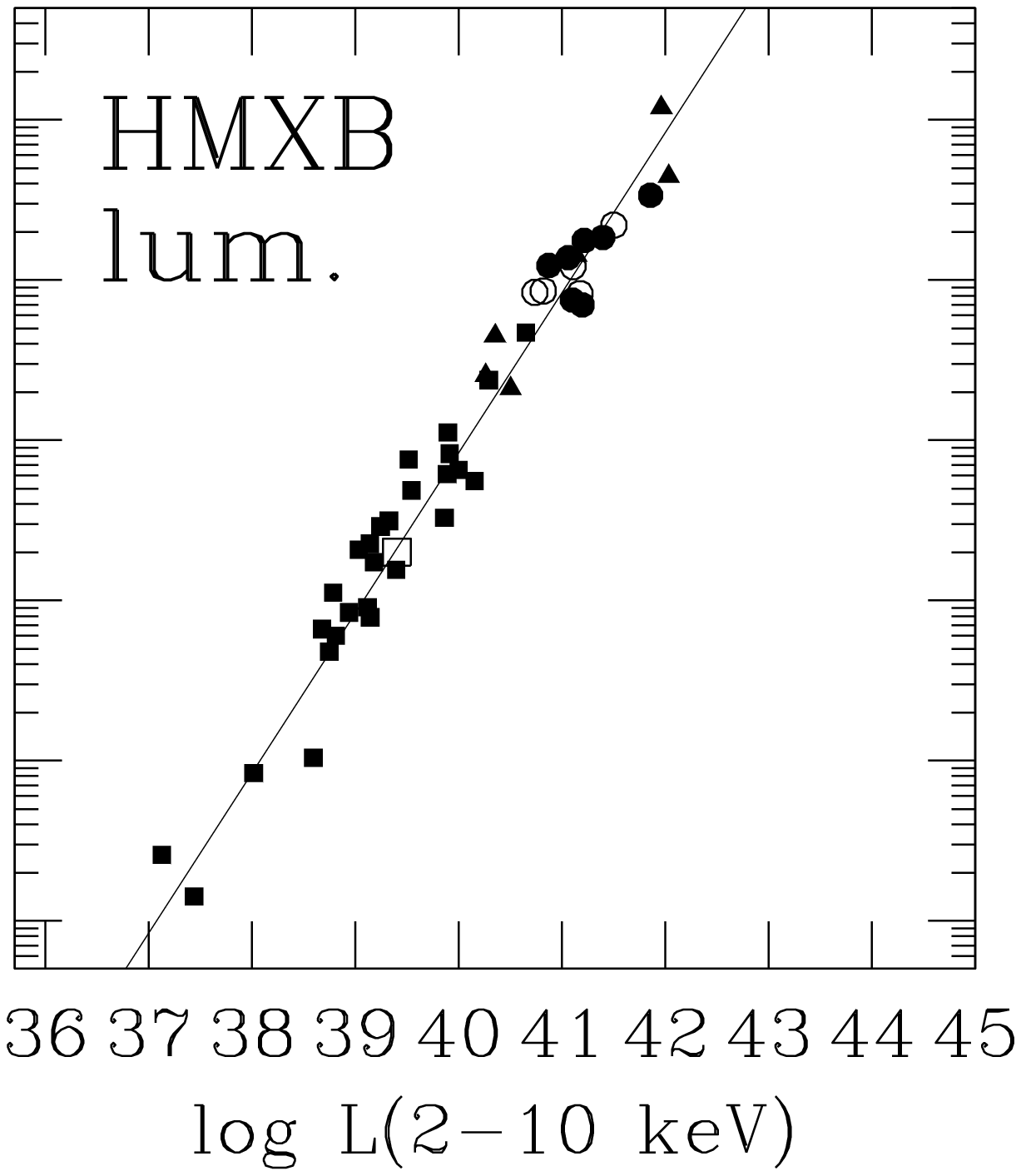}
\caption{The SFR versus 2-10 keV luminosity relations, using the total luminosity 
({\it left}) and the HMXB luminosity ({\it right}). The SFR has been computed from 
the cirrus-subtracted FIR luminosities according to eq.(1) [except for the HDFN 
sample (filled triangles) where radio data have been used; only X-ray detections 
have been used, see Table 3]. Symbols are as in Fig.4; the large empty square 
represents the Milky Way (SFR $\simeq 2\, M_\odot$ yr$^{-1}$, see Matteucci 2002; 
$L_{2-10} \sim 2 \times 10^{39}$ erg s$^{-1}$ and $L_{2-10}^{\rm HMXB} \sim 4 \times 
10^{38}$ erg s$^{-1}$, estimated from Iben et al. 1995a,b and consistent with the 
template spectrum in Fig.1). The solid line in the right panel shows the linear 
relation in eq.(2), whereas the dashed line in the left panel shows the Ranalli et 
al. (2003) relation SFR $= L_{2-10} / (5 \times 10^{39} {\rm erg ~s}^{-1})  M_\odot 
{\rm yr}^{-1}$. }
\end{figure}

\begin{figure}
\vspace{3.3cm}
\includegraphics{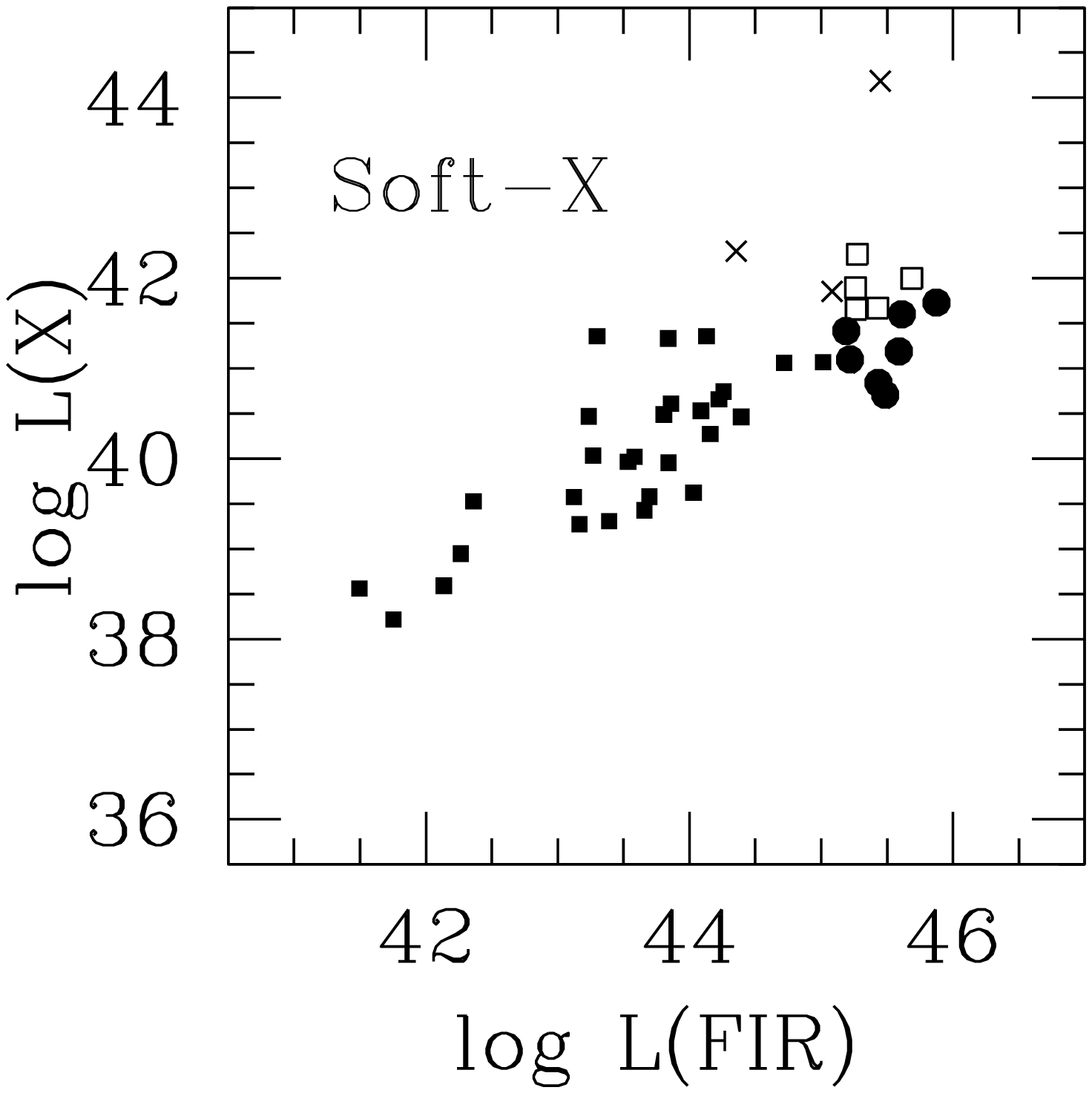}
\hspace{.6cm} 
\includegraphics{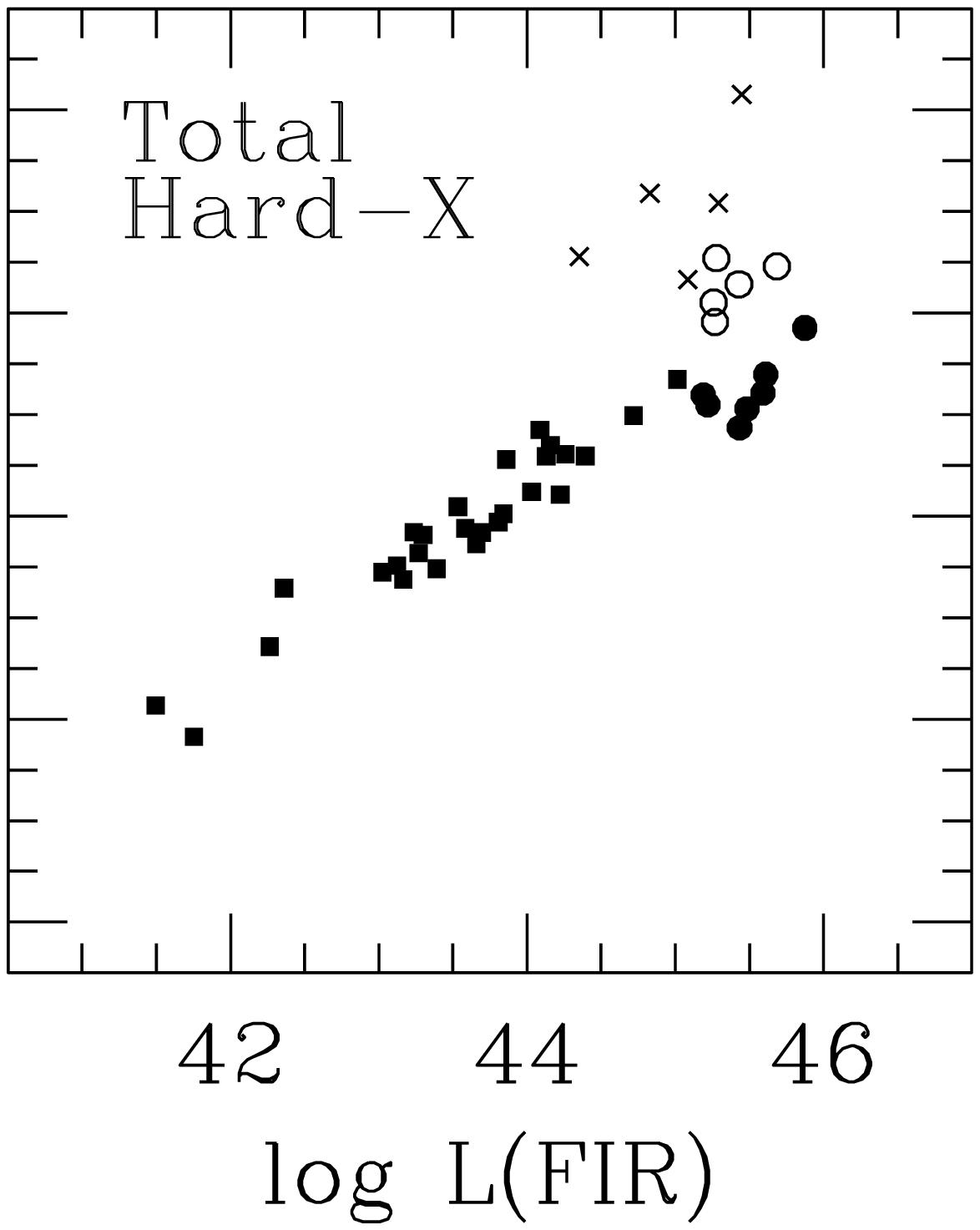}
\hspace{.5cm} 
\includegraphics{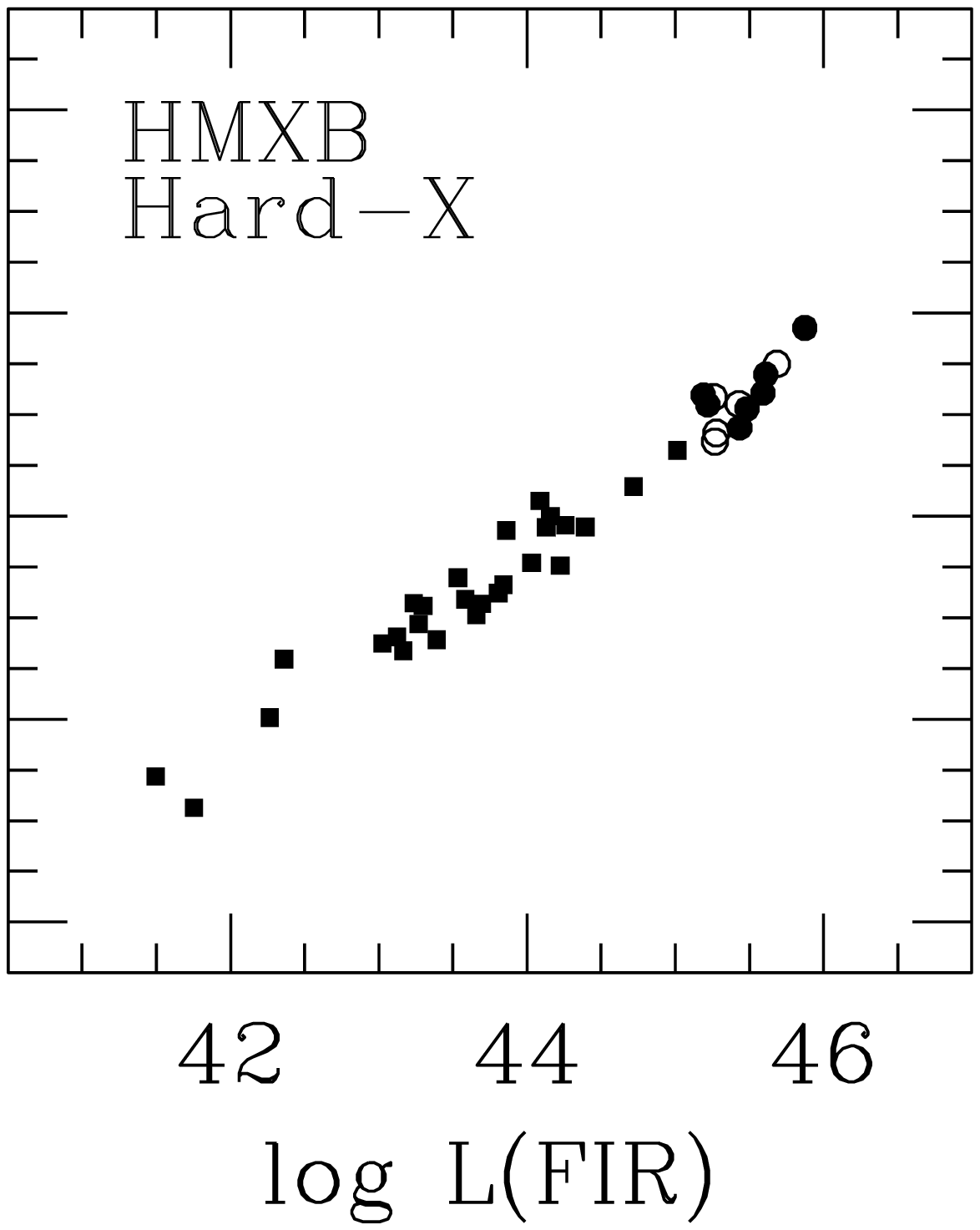}
\caption{ The X-ray--FIR luminosity relation for soft [mostly 0.5-2 keV ({\it left})] and 
hard X-rays [2-10 keV: total emission ({\it center}), and for just HMXB emission ({\it 
right})]. In both panels the FIR luminosities of the SBG sample have been corrected for 
cirrus emission. Symbols are as in Fig.4. }
\end{figure}

\section{Discussion} 

Our main suggestion -- and the central theme of investigation -- has been the notion 
that the 2-10 keV collective emission of HMXBs, $L_{2-10}^{\rm HMXB}$, is a meaningful 
gauge of ongoing galactic SFR. As spelled out in Section 4, given the universality of 
the accretion process onto NS/BH in binaries, the linear dependence of $L_{\rm FIR}$ 
on the SFR, the direct link between the FIR and HMXB emissions, a tight relation is 
predicted between SFR and $L_{2-10}^{\rm HMXB}$. Whether the latter relation is also 
linear depends largely on whether the stellar mass range and the shape of the IMF are 
the same in all galaxies, or vary systematically with the SFR. In the former case we 
expect the relation to be linear, whereas if the IMF becomes progressively more top-heavy 
(i.e., the lower mass cut-off becomes progressively higher) with increasing SFR, a 
non-linear relation is expected. 

As we have mentioned, the use of HMXBs as a galactic SFR estimator was already proposed 
by Grimm et al. (2003) (see also David et al. 1992). Within the general agreement between 
their background picture and ours, the main result of Grimm et al. is, however, quite 
different from ours: their resulting relation is non-linear (SFR $\propto L^{0.6}$) for 
small SFR and low values of $L_{2-10}$ (i.e., SFR $\mincir 4.5 \, M_\odot {\rm yr}^{-1}$ 
and $L_{2-10} \mincir 2.6 \times 10^{40} {\rm erg ~ s}^{-1}$), and linear for higher 
values of the SFR and $L_{2-10}$. Gilfanov et al. (2004) argue that such non-linearity in 
the low-SFR limit may be caused by non-Gaussianity of the probability distribution of the 
integral distribution of discrete sources. We can offer no clear explanation for the 
discrepancy between our result and theirs. One important difference between our approach 
and that of Grimm et al. (2003) is their {\it a priori} minimization of the LMXB 
contribution because -- as they put it -- `owing to the absence of optical identifications 
of a donor star in the X-ray binaries detected by {\it Chandra} in other galaxies, ... 
there is no obvious way to discriminate the contribution of low-mass X-ray binaries'. 
Grimm et al. argue that, given the long evolution timescales of LMXBs, in a galaxy the 
LMXB population should be proportional to the galaxy stellar mass, whereas the HMXB 
population should be proportional to the SFR, so that the relative importance of LMXBs 
should be characterized by the ratio of the stellar mass to the SFR. Using dynamical 
estimates for the stellar masses of galaxies, and SFR derived from a variety of indicators 
(UV, H$\alpha$, FIR, radio, that give a wide range of values of the SFR -- see their Table 
3), Grimm et al. claimed that they selected only galaxies where HMXB emission is expected 
to exceed LMXB emission by a factor of $\magcir$3. For their sample galaxies, consequently, 
Grimm et al. supposed that the plain 2-10 keV luminosity is a measure of the collective 
HMXB emission in that band. In contrast, our procedure involves the identification of the 
HMXB emission from a given SFG by means of a spectral decomposition of the observed X-ray 
spectrum.                                
                                                              
\begin{figure}
\vspace{4.3cm}
\includegraphics{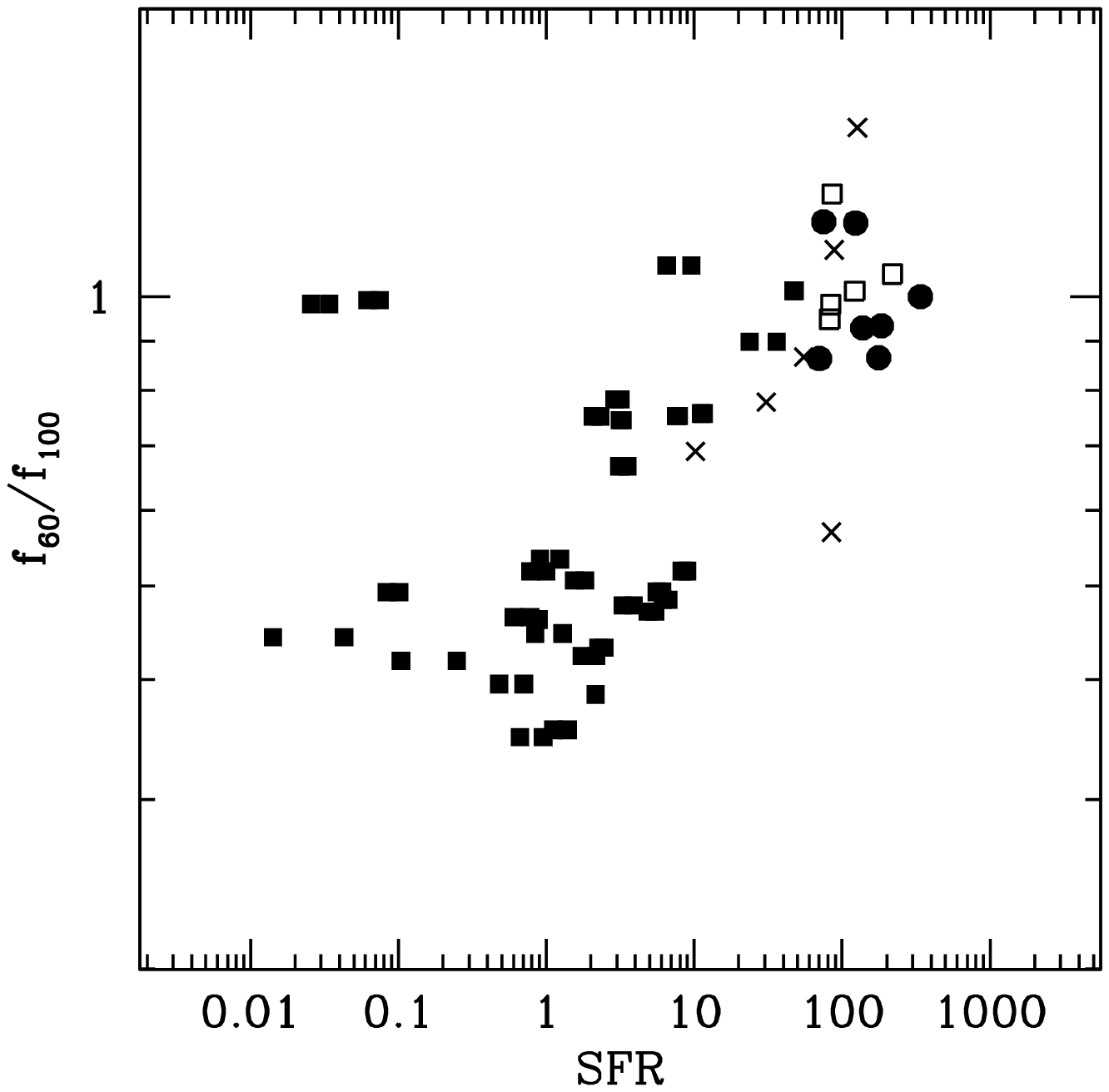}
\hspace{.5cm} 
\includegraphics{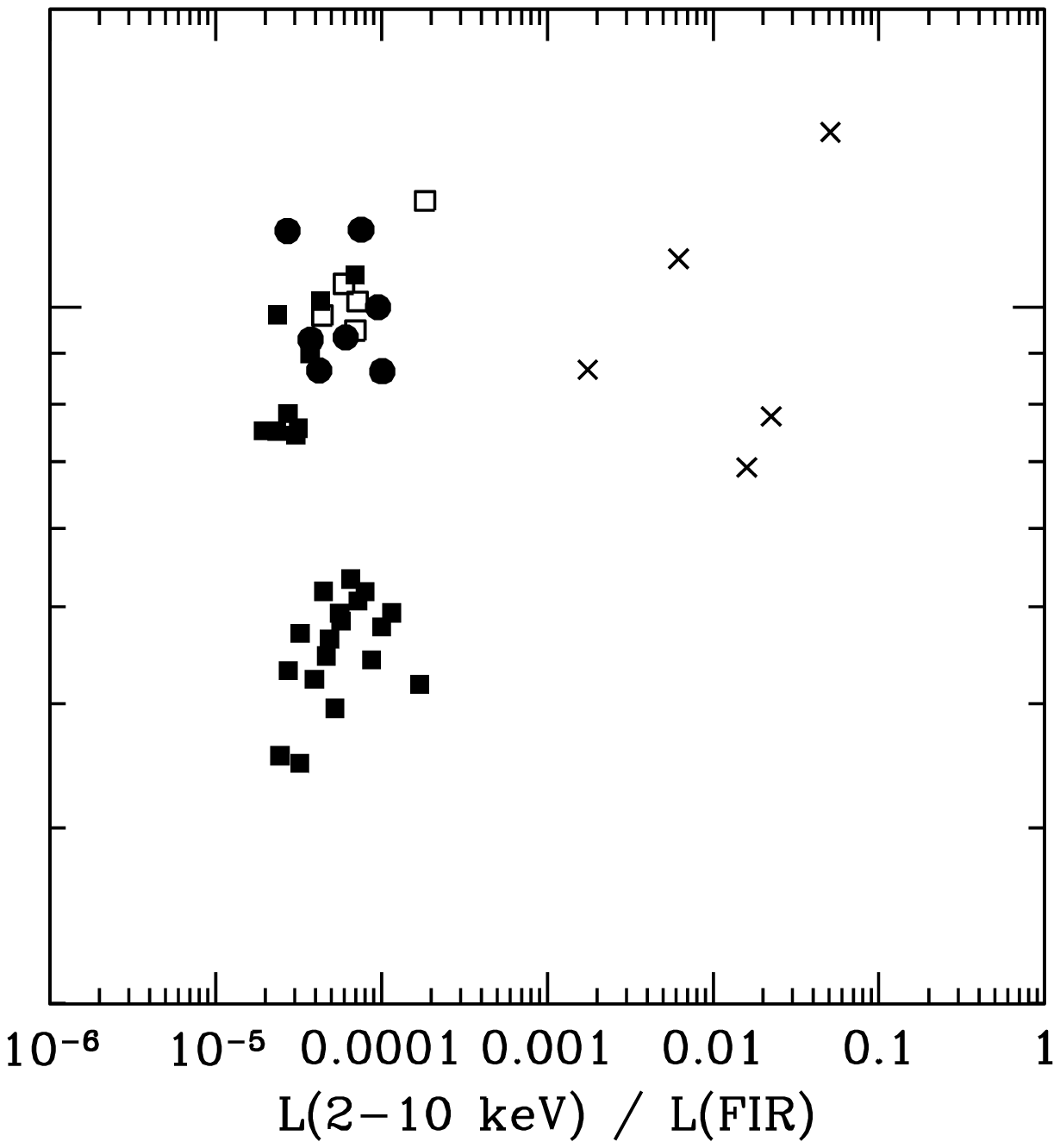}
\caption{ The distribution of 60$\mu$-to-100$\mu$ flux density ratios as a function of the 
SFR ({\it left}) and of the SB 2-10 keV to FIR luminosity ratio ({\it right}). Because 
$f_{60}/f_{100}$ is an approximate measure of the temperature, these plots show that: 
{\it (i)} in spite of a relatively large scatter, a SFR--temperature correlation is 
present in the data, whereby galaxies with higher (lower) SFR tend to have higher (lower) 
dust temperature, $T \sim 50 \, (25) \gr$; and {\it (ii)} the ratio of HMXB 2-10 keV 
luminosity to dust FIR emission remains roughly constant irrespective of the dust 
temperature. Symbols are as in Fig.4. }
\end{figure}

The need in our approach to single out $L_{2-10}^{\rm HMXB}$ rather than $L_{2-10}$ is further 
spurred by the mix of starbursts of widely varying strengths, from low/moderate (local SBGs) to 
extreme (ULIRGs). While the FIR emission is mostly related to current SF in both cases (see 
section 2.1), the 2-10 emission is not in the former case but it is in the latter. Had we 
restricted to the local SBG sample, we would have found SFR $\simeq L_{2-10} / (5 \times 10^{39} 
{\rm erg ~s}^{-1}) M_\odot {\rm yr}^{-1}$ (see Fig.5-{\it left}), in agreement with Ranalli et 
al. (2003). In fact, the 2-10 keV emission from local SBGs has approximately a similar mix of 
SB-related (HMXB) emission and SB-unrelated (LMXB) emission, so that for this sample the primary 
correlation, SFR $\propto L_{2-10}^{\rm HMXB}$, propagates into SFR $\propto L_{2-10}$. Some 
extra scatter plagues the latter relation because the value $f \sim 0.2$ appearing in the equality 
$L_{2-10}^{\rm HMXB} = f \, L_{2-10}$ most likely has some appreciable scatter within the sample. 
Only when the SB-dominated ULIRGs and HDFN galaxies are introduced into the plot and are used as 
calibrators, does the linearity of the SBG SFR--$L_{2-10}$ relation break down (see Fig.5-{\it 
left}). The linearity of the relation, extended to the whole luminosity range, is re-established 
by using the SB-related luminosity $L_{2-10}^{\rm HMXB}$ (see Fig.5-{\it right}). [Of course, the 
calibration of our relation is lower than that of Ranalli et al.'s relation by the factor $f = 
0.2$.]

One key assumption underlying our SFR--$L_{2-10}$ relation is that the FIR luminosities -- from 
which the SFR values are computed -- are really due to the SB and are not contaminated by other 
contributions in any important way. In order to check this assumption, we convert the observed 
$f_{60}/f_{100}$ ratios to dust temperatures using the table in Helou et al. (1988). The 
distribution of 60$\mu$-to-100$\mu$ flux density ratios (see Fig.7-{\it left}) implies a range 
of estimated dust temperatures
	\footnote{Because galaxies have multiple emission components with widely
	varying parameters, the ratio $f_{60}/f_{100}$ cannot be expected to 
	yield a precise temperature. The uncertainty affecting dust temperature 
	estimates is largely intrinsic, and can be traced back to substantial 
	uncertainties that are essentially of astrophysical origin (e.g., the 
	frequency dependence of the dust emissivity, e.g., Helou et al. 1988).  } 
$25 \gr \mincir T \mincir 50 \gr$ (if the dust emissivity is proportional to $\nu^{-1}$), the 
lower bound typically referring to more quiescent (i.e., 'normal') galaxies (e.g., NGC~628) and 
the upper bound to more actively star forming objects (e.g., the well-known SB-dominated galaxy 
Arp~220; Silva et al. 1998). 

This range of values seems realistic for the samples of SFGs considered in this paper, thus 
confirming the soundness of the assumption that our FIR luminosities are not significantly 
contaminated by emission unrelated to ongoing SF activity, and verifying earlier suggestions 
that galaxies with higher SFRs have higher dust temperatures (e.g., David et al. 1992; but 
there may be exceptions to this trend, see Fig.7-{\it left}). Notice also that the ratio of 
the two SB-related luminosities, $L_{2-10}^{\rm HMXB}/L_{\rm FIR}$, remains approximately 
constant when moving from lower temperatures (normal and starburst galaxies) to higher 
temperatures (ULIRGs), log$\, (L_{2-10}/L_{\rm FIR}) \simeq -4.3 \pm 0.3$ (see Fig.7-{\it 
right}). In particular, note that, when properly accounting for LMXB and AGN emission, SBGs 
and ULIRGs imply an X-to-FIR ratio consistent with that for the pure SBs arguably represented 
by the AGN-free ULIRGs. 

In the simple scenario, in which the formation of OB stars and of accreting NSs/BHs during episodes 
of active SF activity only depends on local conditions, the same FIR--X-ray emission relation is 
expected to hold for local ($z \mincir 0.1$) and distant ($z \sim 1$) star-forming galaxies. Indeed, 
the sample of $z \sim 1$ HDFN galaxies we have used does follow the SFR-$L_{\rm x}$ relation 
established in local galaxies. It should be emphasized that distant galaxies are in a substantially 
active SF phase: their 2-10 keV emission must then be HMXB-dominated, if no central AGN is 
contaminating the emission. If an independent estimate of the SFR is also available (e.g., from deep 
radio observations), the location of the galaxy in the SFR-$L_{2-10}$ plane can be used to evaluate 
any excess X-ray luminosity which, if detected, would presumably imply the presence of an AGN. For 
example, if according to eq.(2) a distant galaxy were very overluminous for its SFR (e.g., if $L_{2-
10} \sim 10^{43}$ erg s$^{-1}$ for SFR $\sim 100$ $M_\odot$ yr$^{-1}$), we might suspect a significant 
AGN contribution to the observed 2-10 keV emission (evolutionary considerations rule out significant 
LMXB emission). This method, based on joint X-ray and (say) radio observations, could be a very 
efficient tool to detect AGNs that are buried deeply (i.e., absorbed through $N_{\rm H} > 10^{23}$ 
cm$^{-2}$) in the nuclei of distant SFGs and are not detectable in other bands [e.g., the exemplary 
cases of NGC~6240 (Vignati et al. 1999), NGC~4945 (Guainazzi et al. 2000), Arp~299 (Della Ceca et al. 
2000)]. We therefore caution against using $L_{2-10}$ as a gauge of the SFR.

Finally, from Fig.6-({\it left}) we see that the emission of SFGs at $\sim$1 keV is also an indicator 
of the SFR, although the scatter of the SFR$-L_{\rm x}$ relation is higher in the soft band than in 
the hard band (see Fig.6, where the equivalent $L_{\rm FIR}$--$L_{\rm x}$ relation is shown). Part of 
the scatter seen in Fig.6-({\it left}) is due to observational errors -- such as uncertainties in 
determining the absorption and metallicity of the soft thermal plasma, or occasional differences in 
the definitions of the soft spectral bands -- but most of it is probably intrinsic. In fact, the soft 
component which is systematically observed in the spectra of SFGs is interpreted as sub-keV thermal 
emission originating from SN-powered outgoing galactic winds (e.g., Dahlem et al. 1998; Franceschini 
et al. 2003). Spatially resolved data (Strickland et al. 2000) and simulations (Strickland \& Stevens 
2000) have shown that this emission occurs at the boundary between the hot, tenuous wind fluid proper 
and the cool, denser ISM where conditions are optimal for thermal emission of X-rays. This implies 
that only a small fraction of the wind mass is involved in X-ray emission, and that local conditions 
(i.e., density, clumpiness, and chemical composition of the ISM) crucially determine the level of this 
emission. The small mass fraction of thermally X-ray emitting galactic gas is subject to a large scatter 
among galaxies. These considerations suggest that most of the scatter observed in Fig.6-({\it left}) 
is probably intrinsic. Indeed, the pure {\it thermal} emission does not generally correlate with $L_{\rm 
FIR}$ (see Fig.6c of Franceschini et al. 2003), implying that galactic-wind emission is {\it not} a SFR 
indicator (in spite of winds being an immediate SB outcome). Then, we suggest that the observed $L_{0.5-
2} - L_{\rm FIR}$ correlation is not primary and mainly due to SB-powered galactic winds in the SFGs, 
but that it is probably induced by the portion of the hard spectral component of SFGs showing up in the 
0.5-2 keV band. Notice how the $L_{\rm x} - L_{\rm FIR}$ relation improves when going from left to right 
in Fig.6, i.e. when the reference X-ray luminosity changes from 'total soft' through 'total hard' to 
'HMXB hard'.

\section{Conclusion}

Current limited ability to spectrally decompose the 2-10 keV emission of SFGs is 
nonetheless indicative of the feasibility of using the main HMXB component as 
an estimator of the galactic SFR. The total 2-10 keV luminosity is not a proper SFR 
indicator since for low values of the SFR -- in normal and moderately-starbursting 
galaxies -- it is substantially affected by the emission of LMXBs, while for high 
values of the SFR -- in ULIRGs -- it usually includes significant emission from 
obscured AGNs.

The use of 
an additional sample of distant galaxies from the {\it Hubble} Deep Field North, for 
whom there is no X-ray spectral information, but whose SFR can be estimated from 
deep radio data, led us to to conjecture that the SFR-$L_{\rm x}$ relation that 
we deduced for local galaxies might also hold for distant, $z \sim 1$, galaxies, 
if the 2-10 keV emission of HDFN galaxies is dominated by HMXBs.

The SFR--$L_{2-10}^{\rm HMXB}$ relation may be universal across galaxy types and 
redshifts. This may not be so surprising, if our understanding of the processes 
involved is basically correct. However, its linearity and calibration, both in the 
local and in the distant Universe, are major issues that remain to be settled. 
Resolved high energy long-exposure {\it XMM}-Newton spectra will be needed to afford 
detailed modelling of the integrated emission, while high-spatial-resolution {\it 
Chandra} imaging will be needed to construct the luminosity functions of HMXBs 
Achieving mutually consistent results from the spectral decomposition and from the 
integration of the HMXB luminosity function would lead to a definite determination 
of $L_{2-10}^{\rm HMXB}$. 

Once the universality of the SFR--$L_{2-10}^{\rm HMXB}$ relation has been cleary 
established, it will be possible to use the relation as a tool to unveil the 
presence of deeply absorbed AGNs lurking in the centers of distant galaxies. If, 
for its (independently estimated) SFR, the 2-10 keV luminosity of a distant galaxy 
will be measured to be in excess of that predicted based on the SFR--$L_{\rm x}$ 
relation, that galaxy would be suspected to harbor a strongly absorbed AGN. 

Finally, we stress again that in order to build a physically sound X-ray SFR indicator, 
we should use, in principle, a sample of star-forming galaxies whose X-ray spectral 
properties are well known, and for which the SFR estimates are robust. Extracting the 
HMXB 2-10 keV emission is clearly one crucial step. Ideally, high-quality spectra should 
be available and analyzed for each object in the sample. We unfortunately do not yet have 
such data. At moderate- and low-SFR regimes we relied on estimates gained from analyzing 
data for few objects (M~82, M~83, and NGC~253; plus educated guesses from NGC~2146, 
NGC~2903, NGC~3310, NGC~3256), which we then assumed would hold for the whole sample of 
normal and starburst galaxies. In the high-SFR regime we do have suitable spectral analyses 
for the purely SB-powered ULIRGs, but we had to resort again to estimates in the case of 
AGN-contaminated ULIRGs. For the sample of distant HDFN star-forming galaxies we could only 
make (educated) assumptions. The situation is clearly far from ideal. Therefore our results, 
although promising are clearly preliminary. With the ongoing acquisition of spatially 
resolved {\it Chandra} data and spectrally resolved {\it XMM}-Newton data, it should be 
possible to observe galactic regions that are clearly recognized sites of ongoing SF, in 
particular the actively star-forming central regions of starburst galaxies. These smaller 
starbursts, along with those ULIRGs which are not contaminated by AGNs and can be considered 
as giant starbursts, could yield a well defined SFR--$L_{\rm x}$ relation. So far we have 
been able to suggest a continuity of SF properties moving from normal through starburst to 
ultra-luminous IR galaxies (quantified by $L_{2-10}^{\rm HMXB}/L_{\rm FIR} \simeq 5 \times 
10^{-5}$ over about 5 decades in X-ray luminosity). But the SFR--$L_{\rm x}$ relation we have 
obtained is still affected by a fairly large scatter, which would currently limit its 
applicability even in the case of an object with a well resolved HMXB emission. The picture 
is emerging, the details are still missing.

\begin{acknowledgement}

This research has made use of the NASA/IPAC Extragalactic Database (NED) which is 
operated by the Jet Propulsion Laboratory, California Institute of Technology, 
under contract with the National Aeronautics and Space Administration. Partial
financial support to this project was provided by MIUR through contract 2001028932002
and by ASI through contracts I/R/037/01 and I/R/00206202. We acknowledge the HEASARC 
for providing the {\it ASCA} archival data. We thank an anonymous referee for a very
careful and critical reading of the manuscript.

\end{acknowledgement}

\bigskip

\def\ref{\par\noindent\hangindent 20pt} 

\noindent 
{\bf References} 
\vglue 0.2truecm 

\ref{\small Alexander D.M., Aussel H., Bauer F.E., et al. 2002, ApJ, 568, L85}
\ref{\small Bauer F.E., Alexander D.M., Brandt W.N., et al. 2002, AJ, 124, 2351}
\ref{\small Bauer F.E., Brandt W.N. \& Lehmer B. 2003, AJ, 126, 2797}
\ref{\small Bauer F.E., Brandt W.N., Sambruna R.M., et al. 2001, ApJ, 122, 182}
\ref{\small Bernl\"ohr K. 1993, A\&A, 270, 20}
\ref{\small Blandford R.D. \& Ostriker J.P. 1980, ApJ, 237, 793}
\ref{\small Blustin A.J., Branduardi-Raymont G., Behar E., et al. 2003, A\&A, 403, 481}
\ref{\small Braito V., Franceschini A., Della Ceca R., et al. 2003, A\&A, 398, 107}
\ref{\small Braito V., Della Ceca R., Piconcelli E., et al. 2004, A\&A, in press}
\ref{\small Brandt W.N., Fabian A.C., Takahashi K., et al. 1997, MNRAS, 290, 617}
\ref{\small Burstein D. \& Heiles C. 1982, AJ, 87, 1165}
\ref{\small Cappi M., Persic M., Bassani L., et al. 1999, A\&A, 350, 777}
\ref{\small Christian D.J. \& Swank J.H. 1997, ApJS, 109, 177}
\ref{\small Cohen J.G. 2003, ApJ, 598, 288}
\ref{\small Colbert E.J.M., Heckman T.M., Ptak A.F. \& Strickland D.K. 2003, ApJ, in press 
	(astro-ph/0305476)}
\ref{\small Condon J.J. 1992, ARA\&A, 30, 575}
\ref{\small Dahlem M., Weaver K.A. \& Heckman T.M. 1998, ApJS, 118, 401}
\ref{\small Dahlem M., Parmar A., Oosterbroek T., et al. 2000, ApJ, 538, 555}
\ref{\small David L.P., Jones C. \& Forman W. 1992, ApJ, 388, 82}
\ref{\small Della Ceca R., Griffiths R.E., Heckman T.M. \& MacKenty J.W. 1996, ApJ, 469, 662}
\ref{\small Della Ceca R., Griffiths R.E. \& Heckman T.M. 1997, ApJ, 485, 581}
\ref{\small Della Ceca R., Griffiths R.E., Heckman T.M., et al. 1999, ApJ, 514, 772}
\ref{\small Della Ceca R., Pellegrini S., Bassani L., et al. 2001, A\&A 375, 781}
\ref{\small Della Ceca R., Ballo L., Tavecchio F., et al. 2002, ApJ, 581, L9} 
\ref{\small de Naray P.J., Brandt W.N., Halpern J.P. \& Iwasawa K. 2000, AJ, 119, 612}
\ref{\small Devereux N.A. \& Eales S.A. 1989, ApJ, 340, 708}
\ref{\small Devereux N.A. \& Young J.S. 1991, ApJ, 371, 515}
\ref{\small de Vaucouleurs G., de Vaucouleurs A., Corwin H.G.Jr., et al. 1991, 
	Third Reference Catalogue of Bright Galaxies (New York: Springer Verlag)}
\ref{\small Doane J.S. \& Mathews W.G. 1993, ApJ, 419, 573}
\ref{\small Done C., Madejski G.M., Zycki P.T. \& Greenhill L.J. 2003, ApJ, 588, 763}
\ref{\small Engelbracht C.W. 1997, PhD thesis, Univ. of Arizona}
\ref{\small Fabbiano G. \& White N.E. 2003, in "Compact Stellar X-Ray Sources", eds. 
	W. Lewin \& M. van der Klis (Cambridge University Press) (astro-ph/0307077)}
\ref{\small Fabbiano G., Zezas A. \& Murray S.S. 2001, ApJ, 554, 1035}
\ref{\small Foschini L., Di Cocco G., Ho L.C., et al. 2002, A\&A, 392, 817}
\ref{\small Franceschini A., Braito V., Persic M., et al. 2003, MNRAS, 343, 1181}
\ref{\small Fullmer L. \& Lonsdale C. 1989, "Cataloged Galaxies anf Quasars Observed in the IRAS survey: 
	version 2" (Pasadena: Jet Propulsion Lab)}
\ref{\small Gao Y., Wang Q.D., Appleton PN. \& Lucas R.A. 2003, ApJ, 596, L171}
\ref{\small Garrett M.A., de Bruyn A.G., Giroletti M., et al. 2000, A\&A, 361, L41}
\ref{\small Genzel R., Lutz D., Sturm E., et al. 1998, ApJ, 498, 579}
\ref{\small Gilfanov M., Grimm H.-J. \& Sunyaev R. 2004, MNRAS, 347, L57}
\ref{\small Griffiths R.E., Ptak A., Feigelson E.D., et al. 2000, Science, 250, 1325}
\ref{\small Grimm H.-J., Gilfanov M. \& Sunyaev R. 2003, MNRAS, 339, 793}
\ref{\small Guainazzi M., Matsuoka M., Piro L., et al. 1994, A\&A, 436, L35}
\ref{\small Guainazzi M., Matt G., Brandt W.N., et al. 2000, A\&A, 356, 463}
\ref{\small Helou G., Khan I.R., Malek L. \& Boehmer L. 1988, ApJS, 68, 151}
\ref{\small Helou G., Soifer B.T., Rowan-Robinson M. 1985, ApJ, 298, L7}
\ref{\small Hickson P., Menon K., Palumbo, G. \& Persic M. 1989, ApJ, 341, 679}
\ref{\small Holt S.S., Schlegel E.M., Hwang U. \& Petre R. 2003, ApJ, 588, 792}
\ref{\small Hornschemeier A.E., Brandt W.N., Garmire G.P., et al. 2001, ApJ, 554, 742}
\ref{\small Hornschemeier A.E., Bauer F.E., Alexander D.M., et al. 2003, AJ, 126, 575}
\ref{\small Humphrey P.J., Fabbiano G., Elvis M., et al. 2003, MNRAS, 344, 134}
\ref{\small Hunter D.A., Gillett V.C., Gallagher III J.S., et al. 1986, ApJ, 303, 171}
\ref{\small Iben I.Jr., Tutukov A.V. \& Yungelson L.R. 1995a, ApJS, 100, 217 }
\ref{\small Iben I.Jr., Tutukov A.V. \& Yungelson L.R. 1995b, ApJS, 100, 233}
\ref{\small Inoue A.K., Hirashita H. \& Kamaya H. 2000, PASJ, 52, 539}
\ref{\small Iwasawa K. 1999, MNRAS, 302, 96}
\ref{\small Iwasawa K. \& Comastri A. 1998, MNRAS, 297, 1219}
\ref{\small Iwasawa K., Koyama K., Awaki H., et al. 1993, ApJ, 409, 155}
\ref{\small Iwasawa K., Matt G., Guainazzi M. \& Fabian A.C. 2001, MNRAS, 326, 894}
\ref{\small Kaaret P. 2002, ApJ, 578, 114}
\ref{\small Kaaret P., Prestwich A.H., Zezas A., et al. 2001, MNRAS, 321, L29}
\ref{\small Kennicutt R.C. Jr. 1998, ApJ, 498, 541}
\ref{\small Kilgard R.E., Kaaret P., Krauss M.I., et al. 2002, ApJ, 573, 138}
\ref{\small King A.R. 2003, MNRAS, accepted (astro-ph/0309450)}
\ref{\small Kong A.K.H. 2003, MNRAS, 346, 265)}
\ref{\small Leitherer C. \& Heckman T.M. 1995, ApJS, 96, 9}
\ref{\small Lehnert M. \& Heckman T. 1996, ApJ, 472, 546}
\ref{\small Lira P., Ward M., Zezas A., et al. 2002, MNRAS, 330, 259}
\ref{\small Liu J.-F., Bregman J.N. \& Seitzer P. 2002, ApJ, 580, L31}
\ref{\small Lonsdale Persson C.J. \& Helou G. 1987, ApJ, 314, 513}
\ref{\small Lutz D., Veilleux S. \& Genzel R. 1999, ApJ, 517, L13}
\ref{\small Maccacaro T. \& Perola G.C. 1981, ApJ, 246, L11}
\ref{\small Madau P., Ferguson H.C., Dickinson M.E., et al. 1996, MNRAS, 283, 1388}
\ref{\small Madejski G., Zycki P., Done C., et al. 2000, ApJ, 535, L87}
\ref{\small Maeder A. \& Meynet G. 1989, A\&A, 210, 155}
\ref{\small Martin C.L. \& Kennicutt R.C. 1995, ApJ, 447, 171}
\ref{\small Matteucci F. 2002, lecture given at the XIII Canary Islands Winter School of Astrophysics 
                    'Cosmochemistry: The Melting Pot of Elements' (astro-ph/0203340)
\ref{\small Meurer G.R., Heckman T.M., Lehnert M.D., et al. 1997, AJ, 114, 54}
\ref{\small Mizuno T., Ohbayashi H., Iyomoto N. \& Makishima K. 1998, in IAU Symp. 188 'The Hot Universe' 
	(ed. K.Koyama et al.), 284}
\ref{\small Moran E.C., Lehnert M.D. \& Helfand D.J. 1999, ApJ, 526, 649}
\ref{\small Nandra K., Le T., George I.M., Edelson R.A., et al. 2000, ApJ, 544, 734}
\ref{\small Okada K., Mitsuda K. \& Dotani T. 1997, PASJ, 49, 653}
\ref{\small Perez-Olea D.E. \& Colina L. 1996, ApJ, 468, 191}
\ref{\small Persic M. \& Rephaeli Y. 2002, A\&A, 382, 843}
\ref{\small Persic M. \& Rephaeli Y. 2003, A\&A, 399, 9}
\ref{\small Persic M., Cappi M., Rephaeli Y., et al. 2003, A\&A, submitted}
\ref{\small Prestwich A.H., Irwin J.A., Kilgard R.E., et al. 2002, ApJ, in press (astro-ph/0206127)}
\ref{\small Ptak A.F., Serlemitsos P., Yaqoob T. \& Mushotzky R. 1997, AJ, 113, 1286}
\ref{\small Ptak A., Heckman T., Levenson, N.A., et al. 2003, 592, 782}
\ref{\small Ranalli P., Comastri A. \& Setti, G. 2003, A\&A, 399, 39}
\ref{\small Raymond J.C. \& Smith B.W. 1977, ApJS, 35, 419}
\ref{\small Rephaeli Y. 1979, ApJ, 227, 364}
\ref{\small Rephaeli Y. \& Gruber D.E. 2002, A\&A, 389, 2002}
\ref{\small Rephaeli Y., Gruber D., Persic M. \& McDonald D. 1991, ApJ, 380, L59}
\ref{\small Richards E.A. 2000, ApJ, 533, 611}
\ref{\small Rieke G.H., Loken K., Rieke M.J. \& Tamblyn P. 1993, ApJ 412, 99 }
\ref{\small Roberts T.P., Goad M.R., Ward M.J., et al. 2002b, in "New Visions of
		the X-ray Universe in the XMM-Newton and Chandra Era" (astro-ph/0202017)}
\ref{\small Roberts T.P., Warwick R.S., Ward M.J. \& Murray S.S. 2002a, MNRAS, 337, 677}
\ref{\small Salpeter E.E. 1955, ApJ, ApJ, 121, 161}
\ref{\small Sanders D.B., Mazzarella J.M., Kim D.-C., et al. 2003, AJ, in press (astro-ph/0306263)}
\ref{\small Sansom A.E., Dotani T., Okada K., et al. 1996, MNRAS, 281, 48}
\ref{\small Schlegel D., Finkbeirer D.P. \& Davis M. 1998, ApJ, 500, 525}
\ref{\small Schurch N.J., Roberts T.P. \& Warwick R.S. 2002, MNRAS 335, 241}
\ref{\small Severgnini P., Risaliti G., Marconi A., et al. 2001, A\&A, 368, 44}
\ref{\small Schulz H., Komossa S., Berghoefer Th.W. \& Boer B. 1998, A\&A, 330, 823}
\ref{\small Silva L., Granato G.L., Bressan A. \& Danese L. 1998, ApJ, 509, 103}
\ref{\small Soria R., Pian E. \& Mazzali P.A. 2003, A\&A, in press (astro-ph/0304526)}
\ref{\small Soria R. \& Wu K. 2002, A\&A, 384, 99}
\ref{\small Strickland D.K., Colbert E.J.M., Heckman T.M., et al. 2001, ApJ, 560, 707 }
\ref{\small Strickland D.K., Heckman T.M., Weaver K.A. \& Dahlem M. 2000, AJ, 120, 2965 }
\ref{\small Strickland D.K. \& Stevens, I.R. 2000, MNRAS, 314, 511}
\ref{\small Strohmayer T.E. \& Mushotzky R.F. 2003, ApJ, 586, L61}
\ref{\small Suchov A.A., Balsara D.S., Heckman T.M., Leitherner C. 1994, ApJ, 430, 511}
\ref{\small Swartz D.A., Ghosh K.K., McCollough M.L., et al. 2003, ApJS, 144, 213}
\ref{\small Terashima Y. \& Wilson A.S. 2003, ApJ, accepted (astro-ph/0305563)}
\ref{\small Thompson R.I., Weymann R.J. \& Storrie-Lombardi L.J. 2001, ApJ, 546, 694}
\ref{\small Tully R.B. 1988, Nearby Galaxies Catalog, Cambridge Univ. Press (Cambridge)}
\ref{\small Veilleux S., Kim D.-C. \& Sanders D.B. 1999, ApJ, 522, 139}
\ref{\small Vignati P., Molendi S., Matt G., et al. 1999, A\&A, 349, L57}
\ref{\small Walterbos R.A.M. \& Greenawalt B. 1996, ApJ, 460, 696}
\ref{\small Xia X.Y., Xue S.J., Mao S., et al. 2002, ApJ, 564, 196}
\ref{\small Zezas A., Fabbiano G., Rots A.H. \& Murray S.S. 2002, ApJ, 577, 710}
\ref{\small Zezas A.L., Georgantopoulos I. \& Ward M.J. 1998, MNRAS, 301, 915}
\ref{\small Zezas A., Ward M.J. \& Murray S.S. 2003, ApJ, 594, L31}
\ref{\small White N.E., Swank J.H. \& Holt, S.S. 1983, ApJ, 270, 711}
\bigskip
\bigskip
\bigskip
\bigskip


\centerline{\bf APPENDIX}
\bigskip

In this Appendix we describe the reduction and analysis of the {\it ASCA} and {\it BeppoSAX} 
data for the nearby starburst galaxy M~82. 

\begin{table*}
\caption[] { Spectral fitting results for M~82$^{(a)}$. }
\begin{flushleft}
\begin{tabular}{ l  l  l  l  l  l  l  l  l  l }
\noalign{\smallskip}
\hline
\noalign{\smallskip}
\hline
\hline
 &$\bigg |$ & Wind$^{(b)}$ & & $\bigg |$ Abs.$^{(c)}$ &$\bigg |$ SNR$^{(d)}$ &$\bigg |$ HMXB$^{(e)}$ 
&$\bigg |$ LMXB$^{(f)}$ 
&$\bigg |$ & Fit$^{(g)}$  \\
\hline
\noalign{\smallskip}
 & $\bigg |$ $A_1$ &$A_2$ &$A_3$ &$\bigg |$ $N_{\rm HI}$ &$\bigg |$ $A_4$ &$\bigg |$ $A_5$ &$\bigg |$ $A_6$ &$\bigg |$ $\chi^2_\nu$& DOF \\
\hline
\hline
\\
 SAX & &      &        &         &              &      &   &          &      \\        
\hline
\noalign{\smallskip}
 &$\bigg |$ $11.07^{+2.70}_{-2.70}$ & $2.11^{+0.73}_{-0.73}$E-2 &$3.60^{+0.59}_{-0.59}$E-2 &$\bigg |$ $1.20^{+0.16}_{-0.16}$ 
&$\bigg 
|$ $6.00^{+2.32}_{-2.32}$E-3 &$\bigg |$ $5.04^{+2.46}_{-2.46}$E-4 &$\bigg |$ $7.52^{+0.61}_{-0.61}$E-3 &$\bigg |$ 1.22 & 139 \\
\hline
\hline
\\
 ASCA& &   &        &         &              &      &   &          &         \\        
\hline
\noalign{\smallskip}
 &$\bigg |$ $7.64^{+1.06}_{-1.06}$ & $3.43^{+0.38}_{-0.38}$E-2 &$3.89^{+0.28}_{-0.28}$E-2 &$\bigg |$ $1.27^{+0.07}_{-0.07}$ 
&$\bigg 
|$ $9.04^{+2.13}_{-2.13}$E-3 &$\bigg |$ $7.74^{+4.16}_{-4.16}$E-4 &$\bigg |$ $11.01^{+0.78}_{-0.78}$E-3 &$\bigg |$ 1.41 & 969 \\
\hline
\hline
\\
SAX+\\
ASCA&& &      &        &       &        &                    &   &           \\
\hline
\noalign{\smallskip}
 &$\bigg |$ $8.25^{+1.01}_{-1.01}$ & $3.51^{+0.35}_{-0.35}$E-2 &$3.95^{+0.26}_{-0.26}$E-2 &$\bigg |$ $1.23^{+0.06}_{-0.06}$ 
&$\bigg 
|$ $9.23^{+1.64}_{-1.64}$E-3 &$\bigg |$ $6.49^{+0.03}_{-0.03}$E-4 &$\bigg |$ $11.05^{+0.52}_{-0.52}$E-3 &$\bigg |$ 1.42 & 1114 \\
\hline
\hline

\noalign{\smallskip}
\hline
\end{tabular}
\end{flushleft}
\smallskip

\noindent
$^{(a)}$ The quoted errors represent the estimated $1\, \sigma$ confidence intervals, and are calculated by 
the X-ray spectral-fitting program XSPEC from the derivatives of the fit statistic with respect to the model 
parameters. 

\noindent
$^{(b)}$ Raymond-Smith (1977) plasma models with $kT_1= 0.065$ keV, $kT_2= 0.45$ keV, $kT_3= 0.75$ keV, and  
         $Z_1 = Z_2 = Z_3= 0.1 \, Z_\odot$. The amplitude of the RS plasma model is defined as $A \equiv 
	 10^{-14}/ 4 \pi \bigl((1+z) \, D_{\rm A}\bigr)^2 \int n_{\rm e} n_{\rm H}$, where $D_{\rm A}$ is 
	 the angular size distance to the source (expressed in cm), $ n_{\rm e}$ and $n_{\rm H}$ are the 
	 electron and hydrogen densities (in cm$^{-3}$).

\noindent
$^{(c)}$ HI column density, in excess of the foreground Galactic value, leading to photoelectric absorption 
of emission: units are $10^{22}$ cm$^{-2}$.

\noindent
$^{(d)}$ RS plasma model with $kT=2$ and $Z=Z_\odot$. 

\noindent
$^{(e)}$ Power-law model of the form $A(\epsilon) = A_5 (\epsilon / 1 \, {\rm keV})^{-\Gamma}$ where $A_5$ 
is the amplitude at 1 keV (expressed in photons keV$^{-1}$ cm$^{-2}$ s$^{-1}$) and $\G = 1.2$ is the photon 
index. 

\noindent
$^{(f)}$ Cutoff power-law model of the form $A(\epsilon) = A_6\, (\epsilon / 1 \, {\rm keV})^{-\Gamma}
e^{-\epsilon/\epsilon_{\rm c}}$, where $A_6$ is the amplitude at 1 keV (expressed in photons keV$^{-1}$ 
cm$^{-2}$ s$^{-1}$), $\alpha=1.4$ is the photon index, and $\epsilon_{\rm c}=7.5$ keV is the cutoff energy 
of the exponential cutoff (in keV).

\noindent
$^{(g)}$ All model spectra are absorbed through a foreground Galactic HI column density $N_{\rm HI} = 0.427 
\times 10^{22}$ cm$^{-2}$.
\end{table*}

\bigskip
\centerline{\bf --- Data Reduction ---}
\medskip

\noindent
{\it ASCA data}. 
The available archival spectral data for the galaxies M~82, M~83, NGC~253, 
NGC~2146, NGC~2903, NGC~3256, NGC~3310) were retrieved from the HEASARC for 
analysis. Net spectra for entire observations, typically of duration $20 - 
30$ ks, were obtained, along with response matrices. Provided by the 
project were also the background spectra of the GIS detectors. The SIS 
backgrounds have been computed by accumulating cleaned counts arriving 
in a wide annulus surrounding (but excluding) the source into background 
spectra. Using software provided by the project we then generated response 
matrices for these accumulations
	\footnote{ Note that the derived spectral parameters are 
	sensitive to the background subtraction. Without a proper 
	background subtraction, the hard X-ray spectrum of the 
	source will appear systematically flatter, and the HMXB
	contribution will be overestimated.}. 
Spectral fitting was performed simultaneously on the two SIS and two GIS 
detectors, with a bandpass of 0.4-8 keV for SIS and 0.8-8 keV for GIS. We 
made sure we could closely reproduce the data and the results presented 
for the galaxies in the list by the original authors (M~82, NGC~253: Ptak et 
al. 1997; M~83: Okada et al. 1997; NGC~2146: Della Ceca et al. 1999; NGC~2903: 
Mizuno et al. 1998; NGC~3256: Moran et al. 1999; NGC~3310: Zezas et al. 1998). 

\noindent
{\it BeppoSAX data}. 
The reduction of the {\it BeppoSAX} data for M~82 is described in detail in Cappi 
et al. (1999). Here let it suffice to say that our analysis is restricted to the 
$0.1-4.5$ keV and $1.5-10$ keV energy bands of the LECS and MECS instruments, where 
the response matrices are best calibrated. The extraction region, $4^\prime$, has 
been chosen in order to maximize the S/N ratio. Standard blank-sky files (provided 
by the {\it BeppoSAX} Data Center) were used for background subtraction: the 
background turned out to provide $\sim7\%$ of the total counts at 6 keV. The 
detection of M~82 in the PDS data (nominal band: $13-300$ keV) is statistically 
significant up to $\sim$30 keV. However, some degree of contamination from M~81 
is possible due to the latter's higher flux (by a factor $\sim$2.5) and angular 
proximity (which ensures that about half of its flux is collected). On the other 
hand, the spectral slopes of M~81 and M~82 in the PDS band do differ from each other,
suggesting a tighter constraint on the contamination from M~81. To add uncertainty 
to the issue, it should be recalled that M~81 is known to have varied by a factor 
$\magcir$4 on a timescale of months/years. These considerations led Cappi et al. 
(1999) to a conservatively assume that the PDS data provided only upper limits. In 
our analysis here, although we used all 
(including PDS) data, 
we made sure that the main result of our analysis (i.e., the estimate of the 
fractional 2-10 keV luminosity arising from the HMXB population) is substantially 
solid against including or rejecting the PDS data. 
\bigskip

\centerline{\bf --- Data Analysis ---}
\medskip

Following Persic \& Rephaeli (2002), the model spectrum we have used to analyze the 
M~82 emission comprises a thermal plasma plus the stellar-endproduct populations of 
SNRs, HMXBs, and LMXBs. 

As for the thermal plasma component, we found that the details of the fit at 
$\epsilon \mincir 1$ keV do not appreciably affect the results of the fit (i.e., 
the amplitudes of the stellar endproducts components) at $\epsilon \magcir 2$ keV, 
and vice versa. However, an accurate low-energy fit was required in order to give 
an acceptable value of $\chi^2_\nu$. So we used a multi-phase thermal plasma to fit 
the $< 1$ keV data, allowing both temperature and amplitude to vary, but forcing 
one same chemical abundance for all the phases. In fact, we do have theoretical 
insights 
(e.g., 
Suchkov et al. 1994) and some observational evidence (e.g., Dahlem et al. 1998, 
Strickland et al. 2000) on the existence of multi-phase X-ray emitting thermal 
plasmas in starburst environments, but no well-constrained ranges for the 
temperature and the chemical composition are available other than the rather 
generic (and expected, based on energy considerations) hint that the temperatures 
should be $kT \mincir 1$ keV, and the chemical abundances should be 
$Z \mincir Z_\odot$. Specifically, we found that three plasma phases, described by 
Raymond-Smith (1977) models having temperatures $kT=0.065$, 0.45, 0.75 keV and 
chemical abundance $Z=0.1\, Z_\odot$, provide an acceptable fit to the low-energy 
data. The derived temperatures 
are 
within the range yielded for the X-ray emitting plasma by numerical simulations 
of the interaction between SN-driven winds and the ambient ISM in starbursts 
(Stickland \& Stevens 2000). 
These 
parameter values were determined from an exploration of the parameter space 
$0 \leq kT_1 \leq kT_2 \leq kT_3 \leq 1$, with the constraint that 
$Z_1 \equiv Z_2 \equiv Z_3$. Once these parameters were determined, 
they were kept frozen (in the best-fit procedure) 
while the corresponding amplitudes $A_1$, $A_2$ and $A_3$ were left free to vary. 
As concerns the stellar end-products, we assumed the average spectral shapes 
observed for the corresponding Galactic populations to hold also in the case of 
M~82. 

The results of the fitting, for the separate as well as joint {\it ASCA} and 
{\it BeppoSAX} data sets, are shown in Table 4. Within the errors, the results of 
the separate fits and of the joint fit are mutually consistent.

\end{document}